\documentclass{article}
\usepackage[top=2cm,bottom=3cm,left=2cm,right=2cm]{geometry}
\usepackage{amsmath}
\usepackage{amssymb} 
\usepackage{amsthm}
\usepackage{stmaryrd}
\usepackage{bm}
\usepackage{hyperref}
\usepackage{upgreek}
\usepackage{graphicx}
\usepackage{multirow}

\numberwithin{equation}{section}
\numberwithin{figure}{section}

\def\eq#1{(\ref{eq:#1})}
\def\lineup{\!\!\!\!\!\!\!\!\!\!&&}

\def\d{\partial}
\def\eps{\epsilon}

\def\deg{\mathrm{deg}}

\def\M{{\bf M}}
\def\m{{\bf m}}
\def\Q{{\bf Q}}
\def\n{{\bm \upeta}}
\def\nb{\overline{\bm \upeta}}
\def\b{{\bf b}}
\def\c{{\bf c}}
\def\L{{\bf L}}
\def\mmu{{\bm \upmu}}
\def\l{\lambda}
\def\lb{\overline{\lambda}}
\def\ll{{\bm \uplambda}}
\def\llb{\overline{\bm \uplambda}}
\def\B{{\bf B}}

\def\P{{\bf P}}

\def\A{{\bf A}}
\def\tS{{\tilde{\bf S}}}
\def\S{{\bf S}}
\def\s{{\bf s}}
\def\ssigma{{\bm \sigma}}
\def\G{{\bf \hat{G}}}
\def\q{\tilde{{\bf s}}}
\def\ppi{{\bm \pi}}
\def\Ll{\tilde{{\bm \lambda}}}
\def\p{{\bf p}}

\def\w{\wedge}

\def\[{\big[}
\def\]{\big]}

\begin{document}

\begin{titlepage}
\rightline{\tt LMU-ASC 41/15}
\rightline\today
\begin{center}
\vskip 2cm
\vskip 1.5cm {\large \bf{Ramond Equations of Motion in Superstring Field Theory}}
\vskip 1.0cm
{\large {Theodore Erler\footnote{tchovi@gmail.com}, Sebastian Konopka\footnote{sebastian.konopka@physik.uni-muenchen.de}, Ivo Sachs\footnote{ivo.sachs@physik.uni-muenchen.de}}}
\vskip 1.0cm
{\it {Arnold Sommerfeld Center, Ludwig-Maximilians University}}\\
{\it {Theresienstrasse 37, 80333 Munich, Germany}}

\vskip 2.0cm

{\bf Abstract}
\end{center}

We extend the recently constructed NS superstring field theories in the small Hilbert space to give classical field equations for all superstring theories, including Ramond sectors. We also comment on the realization of supersymmetry in this framework.

\vskip 1.0cm
\noindent 

\noindent

\end{titlepage}

\tableofcontents

\section{Introduction}

The recent construction of NS actions \cite{WittenSS,ClosedSS} has raised the prospect of obtaining a second-quantized, field-theoretic description of all superstring theories. The next step in this programme is to include the Ramond sectors. As is well-known, formulating kinetic terms for the Ramond sector is complicated by the fact that the string field must carry a definite picture \cite{FMS}, which for a holomorphic Ramond state is naturally chosen to be $-1/2$. This is not the right picture to form the usual string field theory kinetic term, 
\begin{equation}\frac{1}{2}\langle \Psi,Q\Psi\rangle,\end{equation}
since in the small Hilbert space the BPZ inner product must act on states whose picture adds up to $-2$. While there are some proposals for circumventing this problem \cite{BerkRamond,Muenster,Michishita,1PIR}, at this time it is not clear what is the most promising way forward. 

Therefore it is worth considering a simpler problem first: namely, constructing classical field equations for all superstring theories, including Ramond sectors. This is the goal of the present paper. With the classical equations of motion, we can
\begin{itemize}
\item compute of tree level amplitudes including Ramond asymptotic states around the perturbative vacuum or any classical solution;
\item investigate the broken and unbroken supersymmetries of classical solutions representing distinct string backgrounds;
\item construct classical solutions in type II closed superstring field theory representing nontrivial Ramond-Ramond backgrounds.
\end{itemize}
The last point is interesting, since Ramond-Ramond backgrounds are quite difficult to describe in the first quantized RNS formalism. While solving the equations of motion of closed string field theory is a tremendously difficult task, it does not appear to be more difficult for Ramond-Ramond backgrounds than other types of background. 

The essential idea behind our construction of the equations of motion is already contained in \cite{WittenSS}. The main new ingredient will be incorporating additional labels associated with multiplication of Ramond states. Like \cite{WittenSS,ClosedSS}, our approach is based on $A_\infty$ and $L_\infty$ algebras, and makes extensive use of associated concepts and notation. We will review the needed apparatus as we go, but for more dedicated discussion see several recent works \cite{WittenSS,ClosedSS,OkWB,WB}.  A different formulation of the equations of motion using the large Hilbert space has already been provided for the open superstring in \cite{BerkRamond} and recently the heterotic string in \cite{Kunitomo1,Kunitomo2}. Our approach has the advantage of describing type II closed superstrings as well, and, once suitable Ramond kinetic terms are formulated, might be generalized to give a classical Batalin-Vilkovisky action. 

\section{Ramond Sector of Open Superstring}
\label{sec:Witten}

In this section we construct the Neveu-Schwarz and Ramond equations of motion for open superstring field theory using Witten's associative star product \cite{Witten}. We will discuss the more general construction based on a non-associative product in the next section. The equations of motion involve two dynamical fields for the NS and R sectors:
\begin{equation}\Phi_\mathrm{N}\in\mathcal{H}_\mathrm{N},\ \ \ \Psi_\mathrm{R}\in\mathcal{H}_\mathrm{R},\end{equation}
where $\mathcal{H}_\mathrm{N}$ and $\mathcal{H}_\mathrm{R}$ are the NS and R open string state spaces, respectively. Both $\Phi_\mathrm{N}$ and $\Psi_\mathrm{R}$ are Grassmann odd and carry ghost number $1$; the NS field $\Phi_\mathrm{N}$ carries picture $-1$ while the Ramond field $\Psi_\mathrm{R}$ carries picture $-1/2$. 

For clarity, let us explain why the Ramond string field is a Grassmann odd object. Any state in the Ramond sector can be built by acting oscillators on the Ramond ground state
\begin{equation}c \,\Theta_{\vec{s}}\,e^{-\phi/2}(0)|0\rangle,\label{eq:Rgs}\end{equation}
where $\Theta_{\vec{s}}$ denotes the spin field \cite{FMS}\footnote{Our conventions concerning the Ramond sector, spinors and gamma matrices follows \cite{Polchinski}.}
\begin{equation}\Theta_{\vec{s}}(z) = \exp\left[i\sum_{a=0}^{4} s_a H_a\right](z),\ \ \ \ \vec{s} = (s_0,s_1,s_2,s_3,s_4),\ \ s_a =\pm\frac{1}{2},\end{equation}
and $H_a, a=0,...,4$ realize the bosonization of the worldsheet fermions through 
\begin{equation}\frac{1}{\sqrt{2}}(\psi^0+\psi^1)=e^{iH_0},\ \ \ \ \ \ \frac{1}{\sqrt{2}}(\psi^{2a}+i\psi^{2a+1})=e^{iH_a}, \ a=1,...,4.\end{equation}
We take the Ramond ground state to be {\it Grassmann even} if it is a positive chirality spinor in the GSO($+$) sector. Then all GSO($+$) projected states built from acting oscillators on \eq{Rgs} will be Grassmann even as well. A Ramond string field is created by taking linear combinations of Ramond states with coefficients which are anticommuting spacetime fields, as is appropriate for fermions. The anticommuting fields anticommute with each other, and in addition we will assume that they anticommute with Grassmann odd worldsheet operators.\footnote{Another convention would assume that anticommuting spacetime fields {\it commute} with anticommuting worldsheet operators. This is closer in spirit to the sign rules of \cite{manifesto},  and in this context the $A_\infty$ and $L_\infty$ algebras we will construct would be equivalently described as {\it super} $A_\infty$ or $L_\infty$ algebras. T.E. thanks U. Schreiber for explaining this convention.} Therefore, in total a Ramond string field must be {\it Grassmann odd}, since it is built from Grassmann even states with Grassmann odd coefficients.

Let us quickly review some notation and conventions which will be essential for our discussion. When discussing $A_\infty$ algebras in open string field theory, it is very useful to use a shifted grading on the open string state space called {\it degree}. The degree of a string field $A$, denoted $\deg(A)$, is defined to be its Grassmann parity plus one (mod $\mathbb{Z}_2$). The dynamical string fields $\Phi_\mathrm{N}$ and $\Psi_\mathrm{R}$ are Grassmann odd, but degree even. Consider a product of $m$ string fields:
\begin{equation}b_m(A_1,...,A_m).\end{equation}
The degree of the product $b_m$, denoted $\deg(b_m)$, is defined to be the degree of its output minus the sum of the degrees of its inputs (mod $\mathbb{Z}_2$). It is useful to think of the product $b_m$ as a linear map from the $m$-fold tensor product of the state space into the state space:
\begin{equation}b_m:\mathcal{H}^{\otimes m} \to\mathcal{H}.\end{equation}
We will write
\begin{equation}b_n(A_1,...,A_m) = b_n(A_1\otimes ... \otimes A_m),\end{equation}
where on the right hand side $b_m$ is regarded as an linear operator acting on tensor products of states. Given a pair of multi-string products $b_m$ and $c_n$, we define a ``commutator"\footnote{Commutators of multi-string products are always graded with respect to degree.}
\begin{equation} 
[b_m,c_n] \equiv b_m\left(\sum_{k=0}^{m-1}\mathbb{I}^{\otimes k}\otimes c_n\otimes \mathbb{I}^{\otimes m-k-1}\right)
  -(-1)^{\deg(b_m)\deg(c_n)} c_n\left(\sum_{k=0}^{n-1}\mathbb{I}^{\otimes k}\otimes b_m\otimes \mathbb{I}^{\otimes n-k-1}\right),\end{equation}
where $\mathbb{I}$ is the identity operator on the state space and
\begin{equation}\mathbb{I}^{\otimes n} =\underbrace{\mathbb{I}\otimes ... \otimes \mathbb{I}}_{n\ \mathrm{times}}\end{equation}  
is the identity operator on $\mathcal{H}^{\otimes n}$. The commutator satisfies the Jacobi identity, and $[b_m,c_n]$ defines an $m+n-1$ string product. Tensor products of operators act in the natural way on tensor products of states. Given two linear maps $b_{k,\ell}$ and $c_{m,n}$ from $\mathcal{H}^{\otimes \ell}\to\mathcal{H}^{\otimes k}$ and $\mathcal{H}^{\otimes n}\to\mathcal{H}^{\otimes m}$, respectively, the tensor product map $b_{k,\ell}\otimes c_{m,n}$ satisfies  
\begin{equation}
b_{k,\ell}\otimes c_{m,n}(A_1\otimes A_2\otimes...\otimes A_{\ell+n}) = (-1)^{\deg(c_{m,n})(\deg(A_1)+...+\deg(A_\ell))}b_{k,\ell}(A_1\otimes...\otimes A_\ell)\otimes c_{m,n}(A_{\ell+1}\otimes...\otimes A_{\ell+n}).
\end{equation}
An $A_\infty$ algebra is defined by a sequence of degree odd multi-string products $d_n,n=1,2,3,...$ which satisfy a hierarchy of identities called {\it $A_\infty$ relations}:
\begin{equation}
[d_1,d_n]+[d_2,d_{n-1}]+...+[d_{n-1},d_2]+[d_n,d_1]=0,\ \ \ \ \ n=1,2,3,...\ .\label{eq:Ainf}
\end{equation}
With these preparations we are ready to discuss the equations of motion. 

Witten's original proposal for open superstring field theory gives the equations of motion \cite{WittenSSFT}
\begin{eqnarray}0\lineup = Q\Phi_\mathrm{N} + X(i) \Phi_\mathrm{N}*\Phi_\mathrm{N} +\Psi_\mathrm{R}*\Psi_\mathrm{R},\\
0\lineup =Q\Psi_\mathrm{R}+X(i)(\Psi_\mathrm{R}*\Phi_\mathrm{N}+\Phi_\mathrm{N}*\Psi_\mathrm{R}),
\end{eqnarray}
where $Q\equiv Q_B$ is the BRST operator, $X(z) = Q\cdot\xi(z)$ is a picture changing operator, and $*$ is the open string star product. As is well known, the Witten theory is singular because of collisions of picture changing operators at the midpoint \cite{Wendt}. We can resolve this problem by spreading the picture changing operators away from the midpoint \cite{OkWB}, which following \cite{WittenSS} we accomplish by postulating that the equations of motion take the form
\begin{eqnarray}0\lineup = Q\Phi_\mathrm{N} + M_2(\Phi_\mathrm{N},\Phi_\mathrm{N}) +m_2(\Psi_\mathrm{R},\Psi_\mathrm{R})+\mathrm{higher\ orders},\label{eq:NEOM}\\
0\lineup =Q\Psi_\mathrm{R} + M_2(\Psi_\mathrm{R},\Phi_\mathrm{N})+M_2(\Phi_\mathrm{N},\Psi_\mathrm{R})+\mathrm{higher\ orders},\label{eq:REOM}
\end{eqnarray}
with higher order terms that we will construct in a moment. The degree odd product $m_2$ is Witten's open string star product with a sign needed to shift the grading from Grassmann parity to degree:
\begin{equation}m_2(A,B) \equiv (-1)^{\deg(A)}A*B.\end{equation}
The degree odd product $M_2$ must carry picture $+1$ and takes the form
\begin{equation}M_2(A,B) = \frac{1}{3}\Big(Xm_2(A,B) + m_2(XA,B) +m_2(A,XB)\Big)\label{eq:M2}\end{equation}
where $X$ is a BPZ even charge of the picture changing operator:
\begin{equation}X\equiv Q\cdot \xi,\ \ \ \ \xi\equiv\oint_{|z|=1}\frac{dz}{2\pi i} f(z) \xi(z).\end{equation}
The function $f(z)$ is holomorphic in the vicinity of the unit circle and is defined so that $\xi$ is BPZ even and anticommutes with the eta zero mode $\eta\equiv \eta_0$ to give 1:
\begin{equation}[\eta,\xi]=1.\end{equation}
An important observation of \cite{WittenSS} is that $M_2$ is BRST exact in the large Hilbert space,
\begin{equation}M_2 = [Q,\mu_2],\end{equation}
and therefore may be formally obtained by an improper field redefinition from a free theory \cite{WB}. Here $\mu_2$ is called the {\it gauge 2-product}\footnote{In \cite{WittenSS} $\mu_2$ was denoted $\overline{M}_2$ and was called the ``dressed 2-product." Our current notation and terminology follows \cite{ClosedSS}.}
\begin{equation}\mu_2(A,B) = \frac{1}{3}\Big(\xi m_2(A,B) - m_2(\xi A,B) -(-1)^{\deg(A)}m_2(A,\xi B)\Big),\label{eq:mu2}\end{equation}
and is degree even. It also satisfies
\begin{equation}m_2 = [\eta,\mu_2].\end{equation}
The choice of $M_2$ in \eq{M2} was dictated in \cite{WittenSS} by cyclicity, i.e. by the assumption that $M_2$ can be derived by varying the cubic vertex of an action. In the current context we are not attempting to construct an action, so cyclicity is not a meaningful requirement. This means that we can in principle make another choice of $M_2$. In fact, there is no reason why the three appearances of $M_2$ in \eq{NEOM} and \eq{REOM} cannot all be chosen to be different products.\footnote{Following the recent suggestion of Sen \cite{1PIR}, one can try to construct a tree-level action with two Ramond fields supplemented by a constraint, in a similar spirit as \cite{Michishita}. In this approach the 2-product of NS states can be chosen to be $M_2$ while the 2-product of an NS and R state should be chosen to be $X m_2$. We did not consider this approach, though it could lead to an interesting refinement of our equations of motion. See also comments in the conclusion.} However, our goal is not necessarily to provide the most general possible form of the equations of motion. We will try, as far as possible, to mimic the construction of the NS sector, which includes some choices which in that context were motivated by cyclicity. 

\subsection{Cubic Order}

The higher order terms in the equations of motion will be defined by a sequence of degree odd multi-string products,
\begin{equation}\tilde{M}_1\equiv Q,\ \tilde{M}_2,\ \tilde{M}_3,\ \tilde{M}_4,\ ...\ ,\end{equation}
which satisfy the relations of an $A_\infty$ algebra. We use the tilde over the products to denote a composite object which appropriately multiplies both NS and R sector states. For example, if $N_1,N_2$ are NS sector string fields and $R_1,R_2$ are R sector string fields, the composite 2-product $\tilde{M}_2$ is defined to satisfy 
\begin{eqnarray}
\tilde{M}_2(N_1,N_2) \lineup \equiv M_2(N_1,N_2),\label{eq:M2NN}\\
\tilde{M}_2(N_1,R_1) \lineup \equiv M_2(N_1,R_1),\\
\tilde{M}_2(R_1,N_1) \lineup \equiv M_2(R_1,N_1),\\
\tilde{M}_2(R_1,R_2) \lineup \equiv m_2(R_1,R_2).\label{eq:M2RR}
\end{eqnarray}
Introducing a composite string field
\begin{equation}\tilde{\Phi} = \Phi_\mathrm{N} + \Psi_\mathrm{R}\in\tilde{\mathcal{H}}\equiv\mathcal{H}_\mathrm{N}\oplus\mathcal{H}_\mathrm{R},\end{equation}
the equations of motion up to second order can be expressed
\begin{equation}0=Q \tilde{\Phi} +\tilde{M}_2(\tilde{\Phi},\tilde{\Phi})+\mathrm{higher\ orders}. \label{eq:WEOM}\end{equation}
Projecting on the NS output (or picture $-1$) gives the equation of motion \eq{NEOM} and projecting on the R output (or picture $-1/2$) gives the equation of motion \eq{REOM}.

Up to cubic order the $A_\infty$ relations are 
\begin{eqnarray}
Q^2 \lineup = 0, \\
\ [Q,\tilde{M}_2] \lineup= 0,\\
\ [Q,\tilde{M}_3] +\frac{1}{2}[\tilde{M}_2,\tilde{M}_2]\lineup=0.
\end{eqnarray}
The first two $A_\infty$ relations are already satisfied since $Q$ is nilpotent and a derivation of both $m_2$ and $M_2$. We will use the third $A_\infty$ relation to determine the composite 3-product $\tilde{M}_3$. First, act the third $A_\infty$ relation on three NS states, or two NS states and one R state. In this case, the commutator $[\tilde{M}_2,\tilde{M}_2]$ reduces to $[M_2,M_2]$, and we can take $\tilde{M}_3=M_3$, where $M_3$ is the 3-product of the NS open superstring field theory found in \cite{WittenSS} (whose form will be reviewed momentarily). Therefore
\begin{eqnarray}
\tilde{M}_3(N_1,N_2,N_3)\lineup  = M_3(N_1,N_2,N_3),\label{eq:M3NNN}\\
\tilde{M}_3(N_1,N_2,R_1)\lineup  = M_3(N_1,N_2,R_1),\\
\tilde{M}_3(N_1,R_1,N_2)\lineup  = M_3(N_1,R_1,N_2),\\
\tilde{M}_3(R_1,N_1,N_2)\lineup  = M_3(R_1,N_1,N_2).\label{eq:M3RNN}
\end{eqnarray}
If there is more than one R input, $\tilde{M}_3$ will take a different form. For example, let us act the third $A_\infty$ relation on three Ramond states:
\begin{eqnarray}
\left([Q,\tilde{M}_3] +\frac{1}{2}[\tilde{M}_2,\tilde{M}_2]\right)R_1\otimes R_2\otimes R_3\lineup = \Big([Q,\tilde{M}_3] +\tilde{M}_2(\tilde{M}_2\otimes\mathbb{I}+\mathbb{I}\otimes \tilde{M}_2)\Big)R_1\otimes R_2\otimes R_3,\nonumber\\
\lineup =\Big([Q,\tilde{M}_3] +M_2(m_2\otimes\mathbb{I}+\mathbb{I}\otimes m_2)\Big)R_1\otimes R_2\otimes R_3,
\end{eqnarray}
where in the second step we acted $\tilde{M}_2$ on the R states to produce $M_2$ and $m_2$. Next we use the fact that $M_2$ is BRST exact in the large Hilbert space:
\begin{equation}
\left([Q,\tilde{M}_3] +\frac{1}{2}[\tilde{M}_2,\tilde{M}_2]\right)R_1\otimes R_2\otimes R_3 = \Big[Q,\Big(\tilde{M}_3 +\mu_2(m_2\otimes\mathbb{I}+\mathbb{I}\otimes m_2)\Big)\Big]R_1\otimes R_2\otimes R_3.
\end{equation}
Since this must be zero, it is natural to identify 
\begin{equation}
\tilde{M}_3(R_1,R_2,R_3) = -\mu_2(m_2\otimes\mathbb{I}+\mathbb{I}\otimes m_2)R_1\otimes R_2\otimes R_3.
\end{equation}
Note that this product is in the small Hilbert space,
\begin{equation}\eta\tilde{M}_3(R_1,R_2,R_3)=0,\end{equation}
since $\eta$ turns $\mu_2$ into $m_2$, and the result vanishes by associativity of $m_2$. Similar considerations determine the remaining 3-products between NS and R states:
\begin{eqnarray}
\tilde{M}_3(N_1,R_1,R_2)\lineup = m_2(\mu_2(N_1,R_1),R_2)-(-1)^{\deg(N_1)}\mu_2(N_1,m_2(R_1,R_2)),\label{eq:M3NRR}\\
\tilde{M}_3(R_1,N_1,R_2)\lineup = m_2(\mu_2(R_1,N_1),R_2)+m_2(R_1,\mu_2(N_1,R_2)),\label{eq:M3RNR}\\
\tilde{M}_3(R_1,R_2,N_1)\lineup = -\mu_2(m_2(R_1,R_2),N_1)+m_2(R_1,\mu_2(R_2,N_1)),\label{eq:M3RRN}\\
\tilde{M}_3(R_1,R_2,R_3)\lineup = -\mu_2(m_2(R_1,R_2),R_3)-(-1)^{\deg(R_1)}\mu_2(R_1,m_2(R_2,R_3)).\label{eq:M3RRR}
\end{eqnarray}
In general, when multiplying $n$ strings there will be $2^n$ formulae representing all ways that NS and R states can multiply. Determining all these formulae seems like a daunting task, but there is a trick to it which we explain in the next subsection.

Before we get to this, however, it is interesting to consider the product of four Ramond states: 
\begin{equation}\tilde{M}_4(R_1,R_2,R_3,R_4).\end{equation}
Since this product would contribute to the NS part of the equations of motion \eq{NEOM}, its ghost number must be $-2$ and its picture number must be $+1$. In fact, this is the first product where the ghost number is more negative than the picture number is positive. It is easy to see that any product built from composing $Q,m_2$ and $\xi$ must satisfy
\begin{equation}\mathrm{ghost\ number}\geq -\mathrm{picture\ number}.\end{equation}
This inequality must be violated for products of four or more Ramond states. Therefore such products potentially present an obstruction to our solution of the $A_\infty$ relations. To see how this problem is avoided, consider the fourth $A_\infty$ relation,
\begin{equation}[Q,\tilde{M}_4]+[\tilde{M_3},\tilde{M_2}] = 0,\end{equation}
acting on four Ramond states:
\begin{eqnarray}
0\lineup=\Big([Q,\tilde{M}_4]+[\tilde{M_3},\tilde{M_2}]\Big)R_1\otimes R_2\otimes R_3\otimes R_4, \nonumber\\
\lineup = \Big([Q,\tilde{M}_4]+\tilde{M_3}(\tilde{M}_2\otimes\mathbb{I}\otimes\mathbb{I} + \mathbb{I}\otimes \tilde{M}_2 \otimes \mathbb{I} + \mathbb{I}\otimes \mathbb{I}\otimes \tilde{M}_2)+\tilde{M}_2(\tilde{M}_3\otimes\mathbb{I}+\mathbb{I}\otimes \tilde{M}_3)\Big)R_1\otimes R_2\otimes R_3\otimes R_4,\nonumber\\
\lineup = \Big([Q,\tilde{M}_4]+\tilde{M_3}(m_2\otimes\mathbb{I}\otimes\mathbb{I} + \mathbb{I}\otimes m_2 \otimes \mathbb{I} + \mathbb{I}\otimes \mathbb{I}\otimes m_2)+m_2(\tilde{M}_3\otimes\mathbb{I}+\mathbb{I}\otimes \tilde{M}_3)\Big)R_1\otimes R_2\otimes R_3\otimes R_4.
\end{eqnarray}
Keeping careful track of the NS and R inputs of $\tilde{M}_3$, this can be further expanded
\begin{eqnarray}
0 \lineup = \Big([Q,\tilde{M}_4]+m_2(\mu_2\otimes\mathbb{I}) (m_2\otimes\mathbb{I}\otimes\mathbb{I})+\mu_2(m_2\otimes m_2)+m_2(\mu_2\otimes\mathbb{I})(\mathbb{I}\otimes m_2 \otimes \mathbb{I})+m_2(\mathbb{I}\otimes\mu_2)(\mathbb{I}\otimes m_2 \otimes \mathbb{I})\nonumber\\
\lineup\ \ \  -\mu_2(m_2\otimes m_2)+m_2(\mathbb{I}\otimes \mu_2)(\mathbb{I}\otimes \mathbb{I}\otimes m_2)-m_2(\mu_2\otimes\mathbb{I})(m_2\otimes\mathbb{I}\otimes\mathbb{I})-m_2(\mu_2\otimes\mathbb{I})(\mathbb{I}\otimes m_2\otimes \mathbb{I})\nonumber\\
\lineup\ \ \ -m_2(\mathbb{I}\otimes \mu_2)(m_2\otimes\mathbb{I}\otimes\mathbb{I})-m_2(\mathbb{I}\otimes \mu_2)(\mathbb{I}\otimes m_2\otimes\mathbb{I})\Big)R_1\otimes R_2\otimes R_3\otimes R_4,\nonumber\\
\lineup = [Q,\tilde{M}_4]R_1\otimes R_2\otimes R_3\otimes R_4.
\end{eqnarray}
Therefore we can simply choose
\begin{equation}\tilde{M}_4(R_1,R_2,R_3,R_4)=0.\label{eq:M4RRRR}\end{equation}
More generally, we claim that all products with four or more Ramond states can be set to zero. Therefore the equations of motion will be cubic in the Ramond string field. 

At first this seems somewhat strange. If the equations of motion have terms which are cubic in the Ramond string field, cyclicity would naturally imply that they should have terms that are quartic in the Ramond string field as well. This is a clear indication that the equations of motion cannot be derived from an action. While this was expected, one might still worry that quartic Ramond terms in the equations of motion are needed to get the correct physics. For example, the quadratic Ramond term $m_2(\Psi_\mathrm{R},\Psi_\mathrm{R})$ is not implied by $A_\infty$ relations or gauge invariance, but is required to incorporate the backreaction of the R field on the NS field. The difference at quartic order is that there is no 4-product of Ramond states at the relevant ghost and picture number which is nontrivial in the small Hilbert space BRST cohomology.  Therefore, any quartic term in the Ramond string field can be removed by field redefinition. As a cross check on our equations of motion, it will be shown in \cite{Konopka} that they imply the correct tree level amplitudes.

\subsection{All Orders}

A key ingredient in constructing the equations of motion at higher order is to realize that multi-string products can be characterized according to their {\it Ramond number}. The Ramond number of a product is defined to be the number of Ramond inputs minus the number of Ramond outputs required for the product to be nonzero: 
\begin{equation}\mathrm{Ramond\ number}=(\mathrm{number\ of\ Ramond \ inputs}) - (\mathrm{number\ of\ Ramond\ outputs}).\end{equation}
Generally, products do not have well-defined Ramond number. A product of Ramond number $N$ has the specific property that it will be nonzero only when multiplying $N$ or $(N+1)$ Ramond states (together with possibly other NS states), in which case it will respectively produce an NS or R state. When Ramond number is defined, we will indicate it by a vertical slash followed by an extra index attached to the product:
\begin{equation}
\setlength{\unitlength}{.25cm}
\begin{picture}(18,6)
\put(0,4.75){$b_{n}|_{N}.$}
\put(1,1.25){\vector(0,1){2.75}}
\put(3,1.25){\oval(4,2)[bl]}
\put(3.25,0){number of inputs}
\put(2.25,3.25){\vector(0,1){.75}}
\put(4.25,3.25){\oval(4,2)[bl]}
\put(4.5,2){Ramond number}
\end{picture}
\end{equation}
If the product is nonzero, its Ramond number must be restricted to the range 
\begin{equation}-1\leq N\leq n,\end{equation}
since the number of Ramond inputs cannot exceed the total number of inputs and the number of Ramond outputs cannot exceed one. While generically multi-string products do not possess well-defined Ramond number, they can always be decomposed into a sum of products which do. To see this, consider the projector
\begin{equation}P_n(N): \tilde{\mathcal{H}}^{\otimes n}\to \tilde{\mathcal{H}}^{\otimes n},\ \ \ P_n(N)^2 = P_n(N),\end{equation}
which selects elements of $\tilde{\mathcal{H}}^{\otimes n}$ which have $N$ Ramond factors. Given an $n$-string product $b_n$, we can define the component at Ramond number $N$:
\begin{equation}b_n|_N =P_1(0)\,b_n\, P_n(N) + P_1(1)\,b_n \,P_n(N+1).\label{eq:Rproj}\end{equation}
Using the resolution of the identity,
\begin{equation}\mathbb{I}^{\otimes n} = \sum_{N=0}^n P_n(N),\end{equation}
it immediately follows that $b_n$ can be expressed as the sum of component products at all Ramond numbers:
\begin{equation}b_n = \sum_{N=-1}^n b_n|_N.\label{eq:bres}\end{equation}
A comment about notation: Generally, we use $b_n|_N$ to denote an $n$-string product of Ramond number $N$, but this does not necessarily mean that $b_n|_N$ is derived from a product $b_n$ after projection to Ramond number $N$. When we do mean this, it should be clear from context. Consider a 1-string product $\mathrm{R}_1$ which acts as the identity on a Ramond state and as zero on an NS state. A product has definite Ramond number $N$ if and only if it satisfies
\begin{equation}\[b_n|_N, \mathrm{R}_{1}\] = N\cdot b_n|_N.\end{equation}
Using the Jacobi identity, this implies that Ramond number is additive when taking commutators of products:
\begin{equation}\[b_m|_{M},c_n|_{N}\]|_{M+N}=\[b_m|_{M},c_n|_{N}\].\end{equation}
Finally, let us mention that the products in the equations of motion always carry {\it even} Ramond number (odd Ramond number components vanish), since picture changing operators do not mix NS and R sector states. Products of odd Ramond number will play a role once we consider supersymmetry in section \ref{sec:SUSY}.

Now let us revisit the results of the previous subsection. The BRST operator has Ramond number zero:
\begin{equation}Q|_0 = Q.\end{equation}
The composite 2-product $\tilde{M}_2$ can be written as the sum of products at Ramond number zero and two. Comparing with equations \eq{M2NN}-\eq{M2RR}, we can apparently write
\begin{equation}\tilde{M}_2 = M_2|_0 + m_2|_2,\label{eq:M2exp}\end{equation}
with the indicated Ramond projection of $M_2$ and $m_2$. Note that $M_2|_0$ can be derived from the Ramond number zero projection of the gauge 2-product $\mu_2$:
\begin{equation}M_2|_0 = \[Q,\mu_2|_0\].\end{equation}
Also,
\begin{equation}m_2|_0 = \[\eta,\mu_2|_0\].\end{equation}
The composite 3-product $\tilde{M}_3$ can likewise be written as the sum of products at Ramond number zero and two: 
\begin{equation}\tilde{M}_3 = M_3|_0 + m'_3|_2.\end{equation}
The Ramond number zero piece corresponds to equations \eq{M3NNN}-\eq{M3RNN}. The Ramond number two piece $m_3'|_2$ is seemingly more complicated, as it must produce four distinct expressions \eq{M3NRR}-\eq{M3RRR} depending on how it multiplies two or three Ramond states. To derive the 3-string products, consider the third $A_\infty$ relation:
\begin{eqnarray}
0 \lineup = [Q,\tilde{M}_3] + \frac{1}{2}[\tilde{M}_2,\tilde{M}_2\],\nonumber\\
\lineup = \[Q,M_3|_0\]+\[Q,m_3'|_2\] + \frac{1}{2}\[M_2|_0,M_2|_0\]+\[M_2|_0,m_2|_2\]+\frac{1}{2}\[m_2|_2,m_2|_2\],\nonumber\\
\lineup =\[Q,M_3|_0\]+\[Q,m_3'|_2\] + \frac{1}{2}\[M_2|_0,M_2|_0\]+\[M_2|_0,m_2|_2\].
\end{eqnarray}
This is equivalent to two independent equations at Ramond number $0$ and $2$:
\begin{eqnarray}
0\lineup = \[Q,M_3|_0\] + \frac{1}{2}\[M_2|_0,M_2|_0\],\\
0\lineup = \[Q,m'_3|_2\] + \[M_2|_0,m_2|_2\].\label{eq:A32}
\end{eqnarray}
The first equation can be solved following \cite{WittenSS}. To review, the solution is
\begin{equation}
M_3|_0 =\frac{1}{2}\Big(\[Q,\mu_3|_0\]+\[M_2|_0,\mu_2|_0\]\Big),\label{eq:M30}
\end{equation}
where $\mu_3|_0$ is the {\it gauge 3-product},
\begin{equation}\mu_3|_0 = \frac{1}{4}\Big(\xi m_3|_0 +m_3|_0(\xi\otimes\mathbb{I}\otimes\mathbb{I} + \mathbb{I}\otimes \xi\otimes\mathbb{I} + \mathbb{I}\otimes\mathbb{I}\otimes \xi)\Big),\end{equation}
and $m_3|_0$ is the {\it bare 3-product},
\begin{equation}
m_3|_0= \[m_2|_0,\mu_2|_0\].
\end{equation}
With these definitions one can show that $M_3|_0$ is in the small Hilbert space:
\begin{equation}\[\eta, M_3|_0\] = 0.\end{equation}
The only difference between $M_3|_0$ and the 3-product of \cite{WittenSS} is the explicit restriction to Ramond number zero. To multiply more than one Ramond state we need $m_3'|_2$. By inspection of the Ramond number two component of the $A_\infty$ relation, we can instantly guess the solution 
\begin{equation}m_3'|_2= \[m_2|_2,\mu_2|_0\].\label{eq:m3p2}\end{equation}
Happily, this simple formula reproduces all four equations \eq{M3NRR}-\eq{M3RRR} for 3-products of two or more Ramond states. To check, for example, we can compute the NRR product:
\begin{eqnarray}
\tilde{M}_3(N_1,R_1,R_2) \lineup = \[m_2|_2,\mu_2|_0\]N_1\otimes R_1\otimes R_2,\nonumber\\
\lineup = \Big(m_2|_2(\mu_2|_0\otimes\mathbb{I}+\mathbb{I}\otimes \mu_2|_0)-\mu_2|_0(m_2|_2\otimes\mathbb{I} + \mathbb{I}\otimes m_2|_2)\Big)
N_1\otimes R_1\otimes R_2,\nonumber\\
\lineup = (m_2(\mu_2\otimes\mathbb{I})-\mu_2(\mathbb{I}\otimes m_2))
N_1\otimes R_1\otimes R_2,\nonumber\\
\lineup = m_2(\mu_2(N_1,R_1),R_2)-(-1)^{\deg(N_1)}\mu_2(N_1,m_2(R_1,R_2)),
\end{eqnarray}
which reproduces \eq{M3NRR}. 

\begin{figure}
\begin{center}
\resizebox{4.5in}{1.2in}{\includegraphics{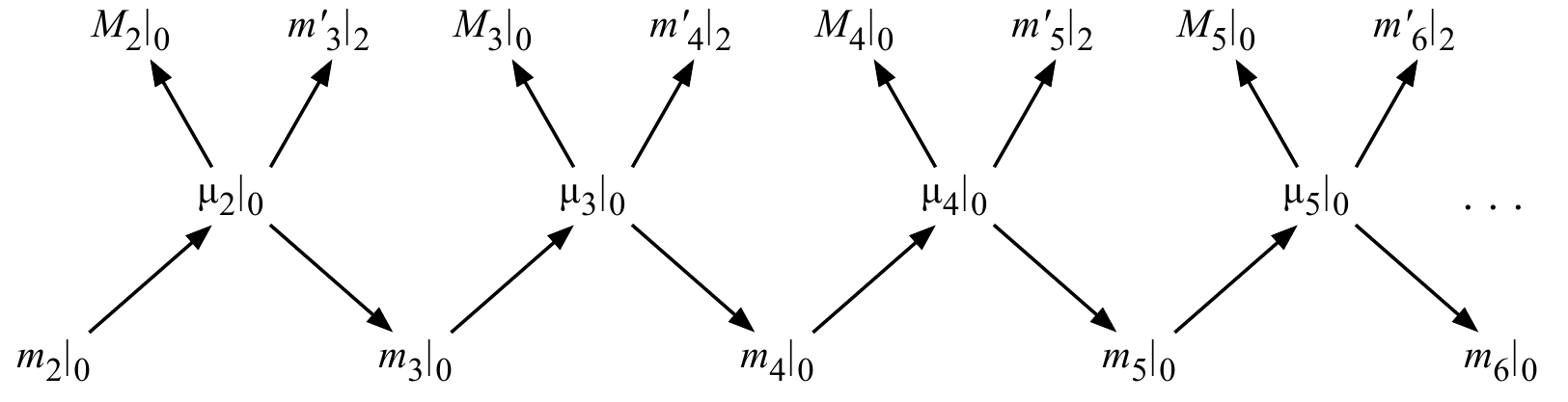}}
\end{center}
\caption{\label{fig:Ramond1} Starting from $m_2|_0$ at the lower left corner, this diagram shows the procedure for constructing all products which appear the NS+R equations of motion using intermediate bare products and gauge products.}
\end{figure}

With some experience from \cite{WittenSS}, it is not difficult to guess the general form of the products to all orders. Let us give the answer first, and then we can prove it. The composite $(n+2)$-string product $\tilde{M}_{n+2}$ can be decomposed
\begin{equation}\tilde{M}_{n+2} = M_{n+2}|_0+m'_{n+2}|_2.\label{eq:tM}\end{equation}
As anticipated before, products with four or more Ramond states can be set to zero. In addition we will need to introduce supplemental {\it bare products} and {\it gauge products}. In total we have four kinds of product: 
\begin{center}
\def\arraystretch{1.5}
\begin{tabular}{rllll}
gauge products & $\mu_{n+2}|_0$: & degree even, & picture\# $= n+1$, & Ramond\#$= 0$,\\
\multirow{2}{*}{ products\ \  $\left\{\displaystyle{\begin{matrix}  \phantom{1}\!\!\!\!\!\!\!\!\\ \phantom{1}\!\!\!\!\!\!\!\!\end{matrix}}\right.$} & $M_{n+1}|_0$: & degree odd, & picture\# $= n$, & Ramond\#$= 0$ \\
 & $m'_{n+2}|_2$: & degree odd, & picture\# $= n$, & Ramond\#$= 2$,\\
bare products & $m_{n+2}|_0$: & degree odd, & picture\# $= n$, & Ramond\#$= 0$,
\end{tabular}
\end{center}
which are determined recursively by the equations:
\begin{eqnarray}
\mu_{n+2}|_0\lineup =\frac{1}{n+3}\left(\xi m_{n+2}|_0-\sum_{k=0}^{n+1}m_{n+2}|_0(\mathbb{I}^{\otimes k}\otimes \xi\otimes\mathbb{I}^{\otimes n+1-k})\right),\label{eq:gaugerec}\\
M_{n+2}|_0\lineup = \frac{1}{n+1}\sum_{k=0}^{n}\[M_{k+1}|_0,\mu_{n-k+2}|_0\],\label{eq:prodrec}\\
m'_{n+3}|_2\lineup = \frac{1}{n+1}\sum_{k=0}^{n}\[m_{k+2}'|_2,\mu_{n-k+2}|_0\],\label{eq:prodprec}\\
m_{n+3}|_0\lineup = \frac{1}{n+1}\sum_{k=0}^{n}\[m_{k+2}|_0,\mu_{n-k+2}|_0\],\label{eq:barerec}
\end{eqnarray}
where 
\begin{equation}M_1|_0\equiv Q,\ \ \ m'_2|_2 \equiv m_2|_2.\end{equation}
The recursive procedure for constructing the products, gauge products, and bare products is illustrated in figure \ref{fig:Ramond1}. Note that these equations are nearly the same as those from \cite{WittenSS} determining the NS open superstring field theory. The only major difference is the appearance of a new set of products $m'_{n+2}|_2$ for multiplying 2 or 3 Ramond states. 

To prove these formulas it is helpful to work with the coalgebra representation of $A_\infty$ algebras. See also \cite{WittenSS,ClosedSS,OkWB,WB}. Consider the tensor algebra generated by taking formal sums of tensor products of string fields: 
\begin{equation}T\mathcal{H} = \mathcal{H}^{\otimes 0}\oplus\mathcal{H}\oplus\mathcal{H}^{\otimes 2}\oplus\mathcal{H}^{\otimes 3}\oplus...\ .\end{equation}
We promote an $m$-string product into a linear operator on the tensor algebra called a {\it coderivation}, which we denote with the same symbol in  boldface:
\begin{equation}b_m\to \b_m.\end{equation}
By definition, the coderivation $\b_m$ acts on the $n$-string component of the tensor algebra as
\begin{equation}
\b_m = \sum_{k=0}^{n-m}\mathbb{I}^{\otimes k}\otimes b_m\otimes \mathbb{I}^{\otimes n-m-k},\ \ \ \ \mathrm{on}\ \mathcal{H}^{\otimes n\geq m},
\end{equation}
and on $\mathcal{H}^{\otimes n< m}$ it gives zero. A commutator of products $[b_m,c_n]$ can also be promoted to a coderivation. This coderivation turns out to be equal to the commutator of the coderivations $\b_m$ and $\c_n$ (graded with respect to degree):
\begin{equation}[b_m,c_n]\to [\b_m,\c_n].\end{equation}
If a sequence of degree odd multi-string products $d_n,n=1,2,3,..$ define an $A_\infty$ algebra, the $A_\infty$ relations can be expressed
\begin{equation}[{\bf d}_1,{\bf d}_n]+[{\bf d}_2,{\bf d}_{n-1}]+...+[{\bf d}_{n-1},{\bf d}_2]+[{\bf d}_n,{\bf d}_1]=0,\ \ \ \ \ n=1,2,3,...\ .\label{eq:codAinf}\end{equation}
The main utility of introducing coderivations is that they can be added to each other. For example, given the products of an $A_\infty$ algebra we can define a coderivation
\begin{equation}{\bf d} = \sum_{n=0}^{\infty} {\bf d}_n.\end{equation}
This single object encapsulates all products of the $A_\infty$ algebra. To get the product $d_n$, we act ${\bf d}$ on the $n$-string component of the tensor algebra and look at the result in the 1-string component of the tensor algebra. Moreover, the $A_\infty$ relations \eq{Ainf} are equivalent to the statement that ${\bf d}$ is nilpotent: 
\begin{equation}[{\bf d},{\bf d}] = 0.\end{equation}
To recover the $A_\infty$ relations \eq{Ainf}, we act this equation on $n$-string component of the tensor algebra and look at the result in the 1-string component of the tensor algebra.

Let us promote the products, gauge products and bare products to coderivations:
\begin{eqnarray}
M_{n+1}|_0\lineup \to\M_{n+1}|_0,\nonumber\\
m'_{n+2}|_2\lineup \to \m'_{n+2}|_2,\nonumber\\
\mu_{n+2}|_0\lineup \to\mmu_{n+2}|_0,\nonumber\\
m_{n+2}|_0\lineup \to\m_{n+2}|_0,
\end{eqnarray}
and define generating functions:
\begin{eqnarray}
\M(t) \lineup = \sum_{n=0}^\infty t^n\M_{n+1}|_0,\label{eq:Mgen}\\
\m'(t)\lineup =\sum_{n=0}^{\infty}t^n \m'_{n+2}|_2,\\
\m(t)\lineup = \sum_{n=0}^\infty t^n\m_{n+2}|_0,\label{eq:mgen}\\
\mmu(t)\lineup = \sum_{n=0}^{\infty}t^n\mmu_{n+2}|_0.\label{eq:mugen}
\end{eqnarray}
Note that 
\begin{eqnarray}
\M(0) \lineup = \Q,\\
\m'(0) \lineup = \m_2|_2,\\
\m(0) \lineup = \m_2|_0.
\end{eqnarray}
Substituting the generating functions and expanding in powers of $t$, it is straightforward to show that equations \eq{gaugerec}-\eq{barerec} are equivalent to: 
\begin{eqnarray}
\frac{d}{dt}\M(t) \lineup = [\M(t),\mmu(t)],\label{eq:Mdiff}\\
\frac{d}{dt}\m'(t) \lineup = [\m'(t),\mmu(t)],\label{eq:mpdiff}\\
\frac{d}{dt}\m(t) \lineup = [\m(t),\mmu(t)],\label{eq:mdiff}\\
\mmu(t) \lineup = \xi\circ \m(t).\label{eq:muxim}
\end{eqnarray}
Here the operation $\xi\circ$ is defined by its action on an $n$-string product
\begin{equation}\xi\circ b_n=\frac{1}{n+1}\left(\xi b_n+(-1)^{\deg(b_n)}b_n\sum_{k=0}^{n-1}\mathbb{I}^{\otimes k}\otimes\xi\otimes\mathbb{I}^{\otimes n-1-k}\right).\end{equation}
This operation defines a homotopy operator for the eta zero mode, in the sense that
\begin{equation}[\n,\xi\circ\b] +\xi\circ[\n,\b] = \b,\end{equation}
where $\b$ is an arbitrary coderivation, and $\n$ is the coderivation representing the $\eta$ zero mode. 

Let ${\bf A}(t)$ or $\B(t)$ stand for $\M(t),\m'(t)$ or $\m(t)$. We have 
\begin{equation}[{\bf A}(0),\B(0)]=0,\end{equation}
since $\Q,\m_2|_0,\m_2|_2$ mutually anticommute. Now note that the differential equations \eq{Mdiff}-\eq{mdiff} imply
\begin{equation}\frac{d}{dt}[{\bf A}(t),\B(t)] = [[{\bf A}(t),\B(t)],\mmu(t)].\end{equation}
Since this equation is homogeneous in $[{\bf A}(t),\B(t)]$, which vanishes at $t=0$, we conclude 
\begin{equation}[{\bf A}(t),\B(t)]=0.\label{eq:MmpmAinf}\end{equation}
In other words, $\M(t),\m'(t)$ and $\m(t)$ are nilpotent and mutually anticommute. Next note that 
\begin{eqnarray}[\n,\A(0)]=0.\end{eqnarray}
since $\Q,\m_2|_0$ and $\m_2|_2$ are in the small Hilbert space. Equations \eq{Mdiff}-\eq{muxim} together with \eq{MmpmAinf} imply
\begin{eqnarray}
\frac{d}{dt}[\n,\A(t)] \lineup = [\n, [\A(t),\mmu(t)]],\nonumber\\
\lineup = [[\n,\A(t)],\mmu(t)]-[\A(t),\m(t)] + [\A(t),\xi\circ[\n,\m(t)]],\\
\lineup =  [[\n,\A(t)],\mmu(t)] + [\A(t),\xi\circ[\n,\m(t)]].
\end{eqnarray}
Suppose $\A(t)=\m(t)$. Then this equation is homogeneous in $[\n,\m(t)]$, and since this vanishes at $t=0$ we conclude $[\n,\m(t)]=0$. Therefore 
\begin{equation}\frac{d}{dt}[\n,\A(t)] = [[\n,\A(t)],\mmu(t)].\end{equation}
Since this equation is homogeneous in $[\n,\A(t)]$, which vanishes at $t=0$, we conclude
\begin{equation}[\n,\A(t)]=0.\end{equation}
In other words, all products and bare products are in the small Hilbert space. Finally, consider the coderivation representing the composite products in the equations of motion
\begin{equation}
\tilde{\M} \equiv \sum_{n=0}^\infty \tilde{\M}_{n+1} = \M(1) + \m'(1).
\end{equation}
The above results immediately imply that 
\begin{equation}[\n,\tilde{\M}] = 0, \ \ \ \ [\tilde{\M},\tilde{\M}] = 0.\end{equation}
The first equation says that the composite products are in the small Hilbert space, and the second equation says that they satisfy $A_\infty$ relations. This completes the construction of the Neveu-Schwarz and Ramond equations of motion for the open superstring based on Witten's open string star product.

\section{Ramond Sector of Open Superstring with Stubs}

In preparation for studying the closed superstring, in this section we provide a more general construction of the open superstring equations of motion which does not require the associativity of Witten's open string star product. Specifically, we build the equations of motion by inserting picture changing operators a set of elementary products at picture zero: 
\begin{equation}M_1^{(0)}\equiv Q,\ \ \ M_2^{(0)},\ \ \ M_3^{(0)},\ \ \ M_4^{(0)},\ \ \ ...\ .\label{eq:bosprod}\end{equation}
We assume that these products have odd degree, live in the small Hilbert space, and satisfy $A_\infty$ relations. For example, we could define the elementary 2-string product $M_2^{(0)}$ by attaching ``stubs" to Witten's open string star product \cite{ClosedSS,Revisited} (hence the title of this section). In the following we will need to introduce a multitude of products with different picture and Ramond numbers. We denote the number of inputs, the picture number and the Ramond number of a product as follows:
\begin{equation}
\setlength{\unitlength}{.25cm}
\begin{picture}(18,9)
\put(2.25,7.25){\vector(0,-1){1}}
\put(4.25,7.25){\oval(4,2)[tl]}
\put(4.5,8){picture number}
\put(0,4.75){$M_{n+1}^{(p)}|_{2r}$,}
\put(2.5,1.25){\vector(0,1){2.75}}
\put(4.5,1.25){\oval(4,2)[bl]}
\put(4.75,0){number of inputs}
\put(4.5,3.25){\vector(0,1){.75}}
\put(6.5,3.25){\oval(4,2)[bl]}
\put(6.75,2){Ramond number}
\end{picture}
\end{equation}
The product $M_{n+1}^{(0)}|_{2r}$ is defined to be the Ramond number $2r$ projection of elementary product $M_{n+1}^{(0)}$. 

The goal is to construct the NS+R equations of motion,
\begin{equation}0=Q\tilde{\Phi}+\tilde{M}_2(\tilde{\Phi},\tilde{\Phi}) + \tilde{M}_3(\tilde{\Phi},\tilde{\Phi},\tilde{\Phi})+\mathrm{higher\ orders},\end{equation}
where $\tilde{\Phi} = \Phi_\mathrm{N} +\Psi_{\mathrm{R}}$ and $\tilde{M}_{n+1}$ are degree odd composite products which appropriately multiply NS and R states. We require that the composite products live in the small Hilbert space and satisfy $A_\infty$ relations. The composite products can be decomposed into a sum of products of definite Ramond and picture number, 
\begin{equation}\tilde{M}_{n+1} = M_{n+1}^{(n)}|_0 + M_{n+1}^{(n-1)}|_2+ M_{n+1}^{(n-2)}|_4+ ... ,\end{equation}
with the sum terminating when the Ramond number exceeds the number of inputs. The picture number is correlated with the Ramond number so that the NS component of the equations of motion will have picture $-1$ and the R component will have picture $-1/2$. The composite 2-product is a sum of two terms:
\begin{equation}\tilde{M}_2 = M_2^{(1)}|_0+ M_2^{(0)}|_2.\end{equation}
The term with Ramond number 2 is the elementary 2-product $M_2^{(0)}$ acting on two Ramond states. The term with Ramond number zero will be defined analogously to \eq{M2}:
\begin{equation}M_2^{(1)}|_0 = \frac{1}{3}\left(XM_2^{(0)}|_0 + M_2^{(0)}|_0(X\otimes\mathbb{I}+\mathbb{I}\otimes X)\right).\label{eq:M2stub}\end{equation} 
We also have 
\begin{eqnarray}
M_2^{(1)}|_0 \lineup =\[Q,\mu_2^{(1)}|_0\],\\
M_2^{(0)}|_0 \lineup =\[\eta,\mu_2^{(1)}|_0\],
\end{eqnarray}
where the gauge product $\mu_2^{(1)}|_0$ is defined 
\begin{equation}\mu_2^{(1)}|_0 = \frac{1}{3}\left(\xi M_2^{(0)}|_0 - M_2^{(0)}|_0(\xi\otimes\mathbb{I}+\mathbb{I}\otimes \xi)\right).\end{equation} 
So far, the equations of motion are precisely the same as in the previous section with the replacement of Witten's associative star product with $M_2^{(0)}$.

At higher orders the non-associativity of $M_2^{(0)}$ begins to play a significant role. The higher order products at zero Ramond number were already described in \cite{ClosedSS}, so let us focus on products with nonzero Ramond number. The composite 3-product is the sum of two terms:
\begin{equation}\tilde{M}_3 = M_3^{(2)}|_0 + M_3^{(1)}|_2.\end{equation}
The $A_\infty$ relations imply that $M_3^{(1)}|_2$ must satisfy
\begin{equation}0=\[Q,M_3^{(1)}|_2\] + \[M_2^{(1)}|_0,M_2^{(0)}|_2\].\end{equation}
Pulling a $Q$ out of this equation we have
\begin{equation}0=\left[Q,M_3^{(1)}|_2 - \[M_2^{(0)}|_2,\mu_2^{(1)}|_0\]\right].\end{equation}
We conclude that $M_3^{(1)}|_2$ satisfies
\begin{equation}M_3^{(1)}|_2 = \[Q,\mu_3^{(1)}|_2\]+ \[M_2^{(0)}|_2,\mu_2^{(1)}|_0\].\end{equation}
Here we introduce a new gauge 3-product $\mu_3^{(1)}|_2$ which is to be defined so that $M_3^{(1)}|_2$ is in the small Hilbert space. In the previous section we could consistently set this product to zero---indeed, in the previous section all gauge products had vanishing Ramond number. If we postulate that 
\begin{equation}\[\eta,\mu_3^{(1)}|_2\] = M_3^{(0)}|_2,\label{eq:nmu312}\end{equation}
where $M_3^{(0)}|_2$ is the Ramond number 2 projection of the elementary 3-product $M_3^{(0)}$, the product $M_3^{(1)}|_2$ will be in the small Hilbert space:
\begin{eqnarray}
\[\eta,M_3^{(1)}|_2] \lineup = -\Big(\[Q,M_3^{(0)}|_2\]+ \[M_2^{(0)}|_2,M_2^{(0)}|_0\]\Big),\nonumber\\
\lineup = - \left.\left(\[Q,M_3^{(0)}\]+ \frac{1}{2}\[M_2^{(0)},M_2^{(0)}\]\right)\right|_{2},\nonumber\\
\lineup = 0,
\end{eqnarray}
as follows from the $A_\infty$ relations for the elementary products. In summary, the product $M_3^{(1)}|_2$ can be constructed by climbing a ``ladder" of products and gauge products, starting from $M_3^{(0)}|_2$:
\begin{eqnarray}
M_3^{(0)}|_2 \lineup = \mathrm{given},\\
\mu_3^{(1)}|_2\lineup = \frac{1}{4}\Big(\xi M_3^{(0)}|_2 - M_3^{(0)}|_2(\xi\otimes\mathbb{I}\otimes\mathbb{I}+\mathbb{I}\otimes \xi\otimes\mathbb{I}+\mathbb{I}\otimes\mathbb{I}\otimes \xi)\Big),\\
M_3^{(1)}|_2 \lineup = \[Q,\mu_3^{(1)}|_2\]+ \[M_2^{(0)}|_2,\mu_2^{(1)}|_0\],
\end{eqnarray}
where the second step inverts \eq{nmu312}. The Ramond number zero piece of the composite 3-product can be found by climbing a similar ladder, as described in \cite{ClosedSS}.

Let us proceed to quartic order. The composite 4-product can be written as the sum of three terms:
\begin{equation}\tilde{M}_4 = M_4^{(3)}|_0 + M_4^{(2)}|_2 + M_4^{(1)}|_4.\end{equation}
In the previous section the Ramond number 4 contribution could be set to zero. Now it will not vanish. Projecting the fourth $A_\infty$ relation,
\begin{equation}0=[Q,\tilde{M}_4] + [\tilde{M}_2,\tilde{M}_3],\end{equation}
onto Ramond number 4 implies
\begin{equation} 0= \[Q,M_4^{(1)}|_4\] + \[M_3^{(1)}|_2,M_2^{(0)}|_2\].\end{equation}
Plugging in the expression for $M_3^{(1)}|_2$ gives
\begin{equation} 0= \[Q,M_4^{(1)}|_4\] + \[\[Q,\mu_3^{(1)}|_2\],M_2^{(0)}|_2\]+\[\[M_2^{(0)}|_2,\mu_2^{(1)}|_0\],M_2^{(0)}|_2\].\end{equation}
The last term is zero since $\[M_2^{(0)}|_2,M_2^{(0)}|_2\]$ would be a 3-product at Ramond number four, which must vanish identically. We therefore conclude that 
\begin{equation}M_4^{(1)}|_4 = \[Q,\mu_4^{(1)}|_4] + \[M_2^{(0)}|_2,\mu_3^{(1)}|_2\],\end{equation}
where the gauge 4-product $\mu_4^{(1)}|_4$ is defined so that $M_4^{(1)}|_4$ is in the small Hilbert space. If we take
\begin{equation}\[\eta,\mu_4^{(1)}|_4\] = M_4^{(0)}|_4,\end{equation}
then
\begin{eqnarray}
\[\eta,M_4^{(1)}|_4\] \lineup = -\Big(\[Q,M_4^{(0)}|_4] + \[M_3^{(0)}|_2,M_2^{(0)}|_2\]\Big),\nonumber\\
\lineup =  -\left.\Big(\[Q,M_4^{(0)}] + \[M_3^{(0)},M_2^{(0)}\]\Big)\right|_4,\nonumber\\
\lineup = 0.
\end{eqnarray}
Therefore, the product $M_4^{(1)}|_4$ does not vanish and can be constructed by climbing a ``ladder" of products and gauge products, starting from $M_4^{(0)}|_4$:
\begin{eqnarray}
M_4^{(0)}|_4 \lineup = \mathrm{given},\\
\mu_4^{(1)}|_4\lineup = \frac{1}{5}\Big(\xi M_4^{(0)}|_4 - M_4^{(0)}|_4(\xi\otimes\mathbb{I}\otimes\mathbb{I}\otimes\mathbb{I}+\mathbb{I}\otimes \xi\otimes\mathbb{I}\otimes\mathbb{I}+\mathbb{I}\otimes\mathbb{I}\otimes \xi\otimes\mathbb{I}+\mathbb{I}\otimes\mathbb{I}\otimes\mathbb{I}\otimes\xi)\Big),\\
M_4^{(1)}|_4 \lineup = \[Q,\mu_4^{(1)}|_4\]+ \[M_2^{(0)}|_2,\mu_3^{(1)}|_2\].
\end{eqnarray}
With sightly more sophisticated use of the Jacobi identity, we can follow this procedure again to construct $M_4^{(2)}|_2$ by climbing the ``ladder:"
\begin{eqnarray}
M_4^{(0)}|_2 \lineup = \mathrm{given},\\
\mu_4^{(1)}|_2\lineup = \frac{2}{5}\Big(\xi M_4^{(0)}|_2 - M_4^{(0)}|_2(\xi\otimes\mathbb{I}\otimes\mathbb{I}\otimes\mathbb{I}+\mathbb{I}\otimes \xi\otimes\mathbb{I}\otimes\mathbb{I}+\mathbb{I}\otimes\mathbb{I}\otimes \xi\otimes\mathbb{I}+\mathbb{I}\otimes\mathbb{I}\otimes\mathbb{I}\otimes\xi)\Big),\\
M_4^{(1)}|_2 \lineup = \[Q,\mu_4^{(1)}|_2\]+ \[M_2^{(0)}|_2,\mu_3^{(1)}|_0\]+ \[M_2^{(0)}|_0,\mu_3^{(1)}|_2\] +  \[M_3^{(0)}|_2,\mu_2^{(1)}|_0\],\\
\mu_4^{(2)}|_2\lineup = \frac{1}{5}\Big(\xi M_4^{(1)}|_2 - M_4^{(1)}|_2(\xi\otimes\mathbb{I}\otimes\mathbb{I}\otimes\mathbb{I}+\mathbb{I}\otimes \xi\otimes\mathbb{I}\otimes\mathbb{I}+\mathbb{I}\otimes\mathbb{I}\otimes \xi\otimes\mathbb{I}+\mathbb{I}\otimes\mathbb{I}\otimes\mathbb{I}\otimes\xi)\Big),\\
M_4^{(2)}|_2 \lineup = \frac{1}{2}\Big(\[Q,\mu_4^{(2)}|_2\]+ \[M_2^{(1)}|_0,\mu_3^{(1)}|_2\]+ \[M_2^{(0)}|_2,\mu_3^{(2)}|_0\] +  \[M_3^{(1)}|_2,\mu_2^{(1)}|_0\]\Big).
\end{eqnarray}
The number of steps in the ladder increases with the number of picture changing operators we need to insert in the product. The Ramond number zero piece of the composite 4-product was already constructed in \cite{ClosedSS}.

\begin{figure}
\begin{center}
\resizebox{6in}{2.2in}{\includegraphics{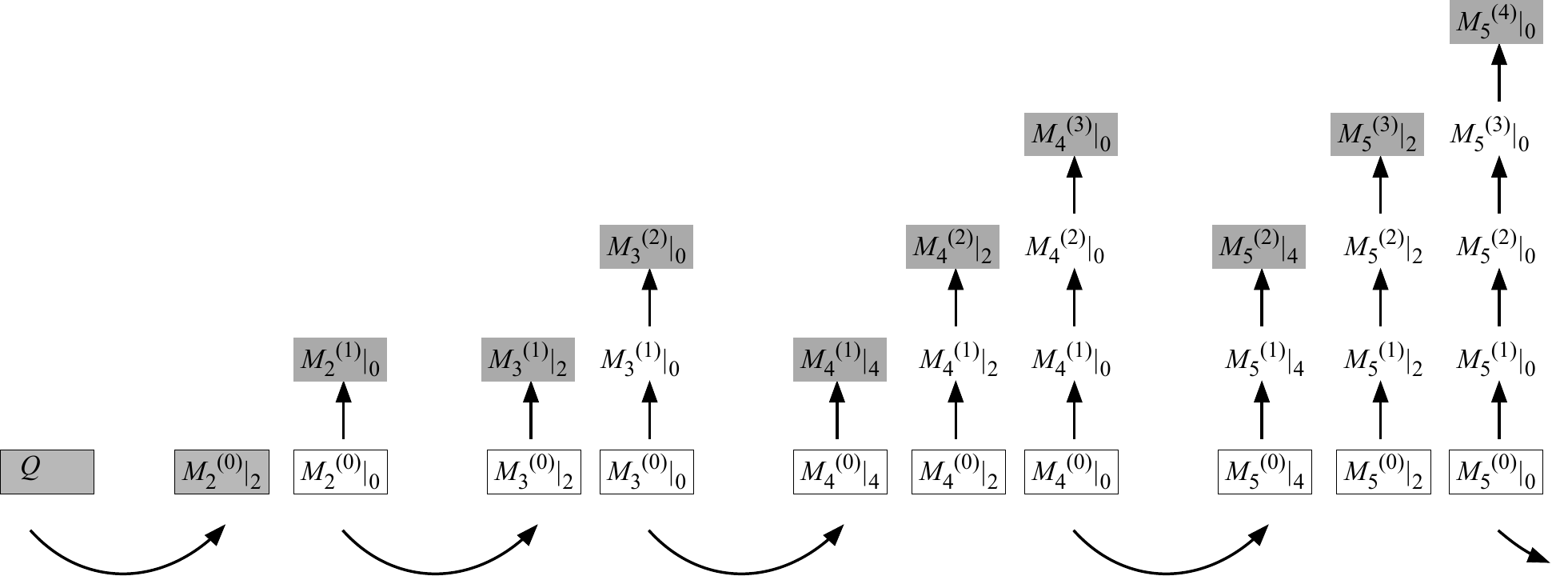}}
\end{center}
\caption{\label{fig:Ramond2} Diagram illustrating the construction of products. The products shaded in grey appear in the equations of motion, and the elementary products at picture zero are boxed. At each order in the string field, the products in the equations of motion are distinguished according to their Ramond number, and at each Ramond number the product is derived by climbing a ``ladder" of products and gauge products starting from an elementary product at picture zero. The ``ladders" are shown above as a stack of products connected by vertical arrows. Each vertical arrow indicates deriving a product with one higher unit of picture from a product of one lower unit of picture by substituting \eq{mustubrec} followed by \eq{Mstubrec}. Climbing each ladder only requires the elementary products and knowledge of products already constructed at lower orders in the string field.}
\end{figure}

Having provided a few examples, let us describe the general construction. We introduce a list of {\it products} and {\it gauge products} as follows:
\begin{eqnarray}
\mathrm{products}:\lineup \ \ \ M_{N+1}^{(p)}|_{2r},\ \ \ \, \mathrm{degree\ odd},\ \ \ \  \nonumber\\
\mathrm{gauge\ products}:\lineup \ \ \ \mu_{N+2}^{(p+1)}|_{2r},\ \ \ \mathrm{degree\ even},
\end{eqnarray} 
where the integers $N,p,r$ take the ranges
\begin{equation}N\geq 0,\ \ \   0\leq r\leq N,\ \ \ 0\leq p\leq N-r. \ \ \ \end{equation}
The restriction on the range of $r$ is a little too generous---many products and gauge products on this list vanish because the Ramond number exceeds the number of inputs. (The first example is $M_3^{(0)}|_4=0$). For convenience we include them on the list anyway. The restriction on the picture number $p$ comes from the fact that the equations of motion require that products have picture number one less than the number of inputs minus half the Ramond number. Since every step on the ladder to construct a product increases the picture, we do not require products whose picture number exceeds this bound. We can alternatively parameterize the list of products and gauge products by three integers $d,p,r$:
\begin{eqnarray}
\mathrm{products}:\lineup \ \ \ M_{d+p+r+1}^{(p)}|_{2r},\ \ \ \, \mathrm{degree\ odd},\nonumber\\
\mathrm{gauge\ products}:\lineup \ \ \ \mu_{d+p+r+2}^{(p+1)}|_{2r},\ \ \ \ \ \mathrm{degree\ even},
\end{eqnarray} 
satisfying 
\begin{equation}d,p,r\geq 0.\end{equation}
The integer $d$ can be interpreted as the picture number {\it deficit}---specifically, the amount of picture that is missing from a product (with a given Ramond number and number of inputs) before it can appear in the equations of motion. The products, gauge products, and bare products introduced in the previous section correspond to a subset of this list:
\begin{eqnarray}
M_{n+1}|_0 \lineup = M_{n+1}^{(n)}|_0,\nonumber\\
m'_{n+2}|_2\lineup = M_{n+2}^{(n)}|_2,\nonumber\\
m_{n+2}|_0 \lineup = M_{n+2}^{(n)}|_0,\nonumber\\
\mu_{n+2}|_0 \lineup = \mu_{n+2}^{(n+1)}|_0.
\end{eqnarray}
If we assume the associativity of Witten's star product, the remaining products and gauge products can be set to zero. Extrapolating from the first few orders, one can guess that the products and gauge products are determined recursively by a pair of equations,
\begin{eqnarray}
\mu_{d+p+r+2}^{(p+1)}|_{2r} \lineup 
= \frac{d+1}{d+p+r+3}\left(\xi M_{d+p+r+2}^{(p)}|_{2r}-M_{d+p+r+2}^{(p)}|_{2r}\sum_{k=0}^{d+p+r+1}\mathbb{I}^{\otimes k}\otimes \xi\otimes \mathbb{I}^{d+p+r+1-k}\right),\label{eq:mustubrec}\\
M_{d+p+r+2}^{(p+1)}|_{2r} \lineup = \frac{1}{p+1}\sum_{p'=0}^p\sum_{d'=0}^d\sum_{r'=0}^r \[ M_{d'+p'+r'+1}^{(p')}|_{2r'},\mu_{d+p+r-d'-p'-r'+2}^{(p-p'+1)}|_{2(r-r')}\].\label{eq:Mstubrec}
\end{eqnarray}
The procedure for constructing the products and gauge products from these equations is illustrated in figure \ref{fig:Ramond2}. 

Now we need to show that that the composite products $\tilde{M}_{n+1}$ which appear in the equations of motion are in the small Hilbert space and satisfy $A_\infty$ relations. We do this by lifting the products and gauge products to coderivations on the tensor algebra, and defining generating functions
\begin{eqnarray}
\M(s,t,u)\lineup =\sum_{p,d,r=0}^{\infty} s^d t^p u^r \M_{d+p+r+1}^{(p)}|_{2r},\\
\mmu(s,t,u)\lineup = \sum_{p,d,r=0}^{\infty}s^d t^p u^r \mmu_{d+p+r+2}^{(p+1)}|_{2r}.
\end{eqnarray}
The parameter $s$ counts the picture deficit, $t$ counts the picture number, and $u$ counts the Ramond number. We can express \eq{mustubrec} and \eq{Mstubrec} in the form of differential equations:
\begin{eqnarray}
\frac{\d}{\d t} \M(s,t,u) \lineup = [\M(s,t,u),\mmu(s,t,u)],\\
\mmu(s,t,u) \lineup = \xi\circ \frac{\d}{\d s}\M(s,t,u).
\end{eqnarray}
These are nearly identical to the differential equations which determine the products of the NS sector \cite{ClosedSS}, the only difference being the label $u$ for Ramond number which just tags along. Next, we claim that
\begin{equation}[\M(s,0,u),\M(s,0,u)] = 0.\label{eq:MstubAinfin}\end{equation}
Unlike in previous examples, this relation is not directly a statement of the $A_\infty$ relations for the elementary products, since the Ramond projections play a nontrivial role. Expanding in powers of $s$ and $u$ gives the formula
\begin{equation}\sum_{r'=0}^r\sum_{d'=0}^d \[M_{d'+r'+1}^{(0)}|_{2r'},M_{d+r-d'-r'+1}^{(0)}|_{2(r-r')}\] = 0.\end{equation}
To prove this, expand the range of summation on the left hand side to write 
\begin{equation}\sum_{r'=0}^r\sum_{d'=0}^d \[M_{d'+r'+1}^{(0)}|_{2r'},M_{d+r-d'-r'+1}^{(0)}|_{2(r-r')}\] = \sum_{r'=0}^r\sum_{d'=-r'}^{d+r-r'} \[M_{d'+r'+1}^{(0)}|_{2r'},M_{d+r-d'-r'+1}^{(0)}|_{2(r-r')}\].\end{equation}
One can check that the additional terms included vanish because the Ramond number of one of the products in the commutator exceeds the number of inputs. Now make the substitution $d'' = d'+r'$, and relabeling $d''\to d'$ we obtain
\begin{eqnarray}\sum_{r'=0}^r\sum_{d'=0}^d \[M_{d'+r'+1}^{(0)}|_{2r'},M_{d+r-d'-r'+1}^{(0)}|_{2(r-r')}\] \lineup 
= \sum_{d'=0}^{d+r}\sum_{r'=0}^r \[M_{d'+1}^{(0)}|_{2r'},M_{d+r-d'+1}^{(0)}|_{2(r-r')}\],\nonumber\\
\lineup 
=\sum_{d'=0}^{d+r} \left.\[M_{d'+1}^{(0)},M_{d+r-d'+1}^{(0)}\]\right|_{2r},\nonumber\\
\lineup = 0,\end{eqnarray}
which vanishes as a consequence of the $A_\infty$ relations for the elementary products. This proves \eq{MstubAinfin}. Next note that 
\begin{equation}\frac{\d}{\d t} [\M(s,t,u),\M(s,t,u)] = [[\M(s,t,u),\M(s,t,u)],\mmu(s,t,u)].\end{equation}
Since this equation is homogeneous in $[\M(s,t,u),\M(s,t,u)]$, which vanishes at $t=0$ as just demonstrated, we conclude
\begin{equation}[\M(s,t,u),\M(s,t,u)]= 0.\end{equation}
Since the elementary products at picture zero are assumed to be in the small Hilbert space, we have
\begin{equation}[\n,\M(s,0,u)]=0.\end{equation}
Next consider the equation
\begin{eqnarray}
\frac{\d}{\d t} [\n,\M(s,t,u)]\lineup = [\n,[\M(s,t,u),\mmu(s,t,u)]],\nonumber\\
\lineup = [[\n,\M(s,t,u)],\mmu(s,t,u)]-\left[\M(s,t,u),\frac{\d}{\d s}\M(s,t,u)\right]+\left[\M(s,t,u),\xi\circ\frac{\d}{\d s}[\n,\M(s,t,u)]\right],\nonumber\\
\lineup = [[\n,\M(s,t,u)],\mmu(s,t,u)]-\frac{1}{2}\frac{\d}{\d s}[\M(s,t,u),\M(s,t,u)]+\left[\M(s,t,u),\xi\circ\frac{\d}{\d s}[\n,\M(s,t,u)]\right],\nonumber\\
\lineup = [[\n,\M(s,t,u)],\mmu(s,t,u)]+\left[\M(s,t,u),\xi\circ\frac{\d}{\d s}[\n,\M(s,t,u)]\right].
\end{eqnarray}
Since this equation is homogeneous in $[\n,\M(s,t,u)]$, which vanishes at $t=0$, we conclude
\begin{equation}[\n,\M(s,t,u)]=0,\end{equation}
which implies that all products are in the small Hilbert space. Finally, consider the coderivation for the composite products which appear in the equations of motion:
\begin{equation}\tilde{\M} = \sum_{n=0}^\infty \tilde{\M}_{n+1} = \M(0,1,1).\end{equation}
The above results imply that 
\begin{equation}[\n,\tilde{\M}]=0,\ \ \ \ [\tilde{\M},\tilde{\M}] = 0,\end{equation}
so the composite products are in the small Hilbert space and satisfy $A_\infty$ relations. This completes the construction of the Neveu-Schwarz and Ramond equations of motion for the open superstring with stubs.

\section{Ramond Sector of Heterotic String}

The generalization of the open superstring with stubs to the heterotic string is straightforward. It requires two steps: First we symmetrize the tensor algebra, which replaces the $A_\infty$ structure of the previous section with an $L_\infty$ structure. Second, we fix the operator $\xi$ to be the zero mode $\xi_0$ for consistency with the level matching and $b_0^{-}$ constraints. Those familiar with \cite{ClosedSS} may need no further explanation, but for completeness we spell it out here. 

Closed string products are naturally commutative---they are symmetric (up to signs) upon interchange of entries. This means that closed string products are naturally understood as linear maps on symmetrized tensor products of closed string states. We denote the symmetrized tensor product with a wedge $\w$. Given string fields $\Phi_1,...,\Phi_n$ in the closed string state space $\mathcal{H}$, their symmetrized tensor product is defined
\begin{equation}\Phi_1\w\Phi_2\w ... \w\Phi_n \equiv \sum_{\sigma} (-1)^{\eps(\sigma)} \Phi_{\sigma(1)}\otimes\Phi_{\sigma(2)}\otimes...\otimes\Phi_{\sigma(n)},
\label{eq:Phiw}
\end{equation}
where the sum is over all permutations $\sigma$ of $1,...,n$ and the sign $(-1)^{\eps(\sigma)}$ is the obvious sign obtained from moving the string fields past each other into the order prescribed by $\sigma$. The commuting/anticommuting property of closed string fields is measured by their {\it degree}. For a closed string field, the degree is taken to be the same as Grassmann parity---not Grassmann parity plus one, as for open string fields. Symmetrized tensor products of $n$ closed string states form a basis for $\mathcal{H}^{\w n}$, and a closed string product $b_n$ is a linear map:
\begin{equation}b_n:\mathcal{H}^{\w n}\to \mathcal{H}.\end{equation}
We write 
\begin{equation}b_n(\Phi_1,...,\Phi_n) = b_n(\Phi_1\w ... \w\Phi_n),\end{equation}
where on the left and side $b_n$ is written as a product and on the right hand side $b_n$ written as an operator on the symmetrized tensor product of states. Given two linear maps $b_{k,\ell}$ and $c_{m,n}$ from $\mathcal{H}^{\w \ell}\to\mathcal{H}^{\w k}$ and $\mathcal{H}^{\w n}\to\mathcal{H}^{\w m}$, respectively, we define a wedge product, 
\begin{equation}b_{k,\ell}\w c_{m,n}:\mathcal{H}^{\w \ell+n}\to\mathcal{H}^{\w k+m},\end{equation}
as follows:
\begin{equation}
b_{k,\ell}\w c_{m,n}(\Phi_1\w ... \Phi_{\ell+n}) = \sum_{\sigma}\,\!' (-1)^{\eps(\sigma)} b_{k,\ell}(\Phi_{\sigma(1)}\w ... \w\Phi_{\sigma(\ell)})\w c_{m,n}(\Phi_{\sigma(\ell+1)}\w ... \w\Phi_{\sigma(n+\ell)}),
\end{equation}
where the sum $\sum_{\sigma}\,\!'$ is over all permutations $\sigma$ of $1,...,n+\ell$ which change the inputs of $b_{k,\ell}$ and $c_{m,n}$. (Permutations which only move around entries of $b_{k,\ell}$ and $c_{m,n}$ produce the same terms, and are only counted once). The sign $(-1)^{\eps(\sigma)}$ is the obvious sign obtained from moving the string fields past each other and $c_{m,n}$ into the order prescribed by $\sigma$. Given two closed string products $b_m$ and $c_n$, we define their commutator
\begin{equation}[b_m,c_n]\equiv b_m(c_n\w\mathbb{I}_{m-1}) - (-1)^{\deg(b_m)\deg(c_n)} c_n(b_m\w\mathbb{I}_{n-1}),\end{equation}
where $\mathbb{I}_n$ is the identity operator on $\mathcal{H}^{\w n}$. The commutator bracket satisfies the Jacobi identity, and $[b_m,c_n]$ is an $(m+n-1)$-closed-string product. An $L_\infty$ algebra is defined by a sequence of degree odd closed string products $d_n,n=1,2,3,...$ which satisfy a hierarchy of identities called $L_\infty$ relations:
\begin{equation}
[d_1,d_n]+[d_2,d_{n-1}]+...+[d_{n-1},d_2]+[d_n,d_1]=0,\ \ \ \ \ n=1,2,3,...\ .
\end{equation}
Finally, we will find it convenient to consider symmetrized tensor algebra
\begin{equation}S\mathcal{H}=\mathcal{H}^{\w 0}\oplus\mathcal{H}\oplus\mathcal{H}^{\w 2}\oplus\mathcal{H}^{\w 3}\oplus...\ .\end{equation}
We promote an $m$-closed-string product to {\it coderivation} on the symmetrized tensor algebra, which we denote with the same symbol in boldface
\begin{equation}b_m\to\b_m.\end{equation}
By definition, $\b_m$ acts on the $n$-string component of the symmetrized tensor algebra as 
\begin{equation}\b_m = b_m\w\mathbb{I}_{n-m} \ \ \ \mathrm{on}\ \mathcal{H}^{\w m\geq n},\end{equation}
and on $\mathcal{H}^{\w m<n}$ it acts as zero. The commutator $[b_m,c_n]$ promotes to a coderivation which is equal to the commutator of coderivations:
\begin{equation}[b_m,c_n] \to [\b_m,\c_n].\end{equation}
It should be emphasized that, as far as our analysis is concerned, the treatment of open string $A_\infty$ algebras and closed string $L_\infty$ algebras is nearly identical. The only difference is that open string products and their commutators act in the appropriate way on tensor products of open string states, while closed string products and their commutators act in the appropriate way on symmetrized tensor products of closed string states. 

Now let us discuss the NS+R equations of motion of the heterotic string. We need two dynamical fields for the NS and R sectors:
\begin{equation}\Phi_{\mathrm{N}}\in \mathcal{H}_{\mathrm{N}},\ \ \ \Psi_{\mathrm{R}}\in \mathcal{H}_{\mathrm{R}},\end{equation}
where $\mathcal{H}_{\mathrm{N}}$ and $\mathcal{H}_{\mathrm{R}}$ are the Neveu-Schwarz and Ramond state spaces of a heterotic string. For the heterotic string, superconformal ghosts and picture only inhabit the left-moving sector.  The NS field $\Phi_{\mathrm{N}}$ should be degree even (and Grassmann even), ghost number 2, and picture number $-1$. The Ramond field $\Psi_{\mathrm{R}}$ should be degree even (and Grassmann even), ghost number 2, and picture number $-1/2$. With the requisite GSO projection, the Ramond field is a linear combination of Grassmann odd states with Grassmann odd coefficients. We define a composite heterotic string field
\begin{equation}\tilde{\Phi}=\Phi_{\mathrm{N}}+\Psi_{\mathrm{R}}\in\tilde{\mathcal{H}}=\mathcal{H}_{\mathrm{N}}\oplus\mathcal{H}_{\mathrm{R}}.\end{equation}
This must satisfy the usual $b_0^{-}$ and level matching constraints:
\begin{equation}b_0^{-}\tilde{\Phi}=0,\ \ \ \ L_0^{-}\tilde{\Phi}=0,\end{equation}
where $b_0^{-} = b_0-\overline{b}_0$ and $L_0^{-} = L_0-\overline{L}_0$ is the difference between left and right-moving zero modes. The NS+R equations of motion of the heterotic string take the form
\begin{equation}0=Q\tilde{\Phi} + \tilde{L}_2(\tilde{\Phi},\tilde{\Phi})+ \tilde{L}_3(\tilde{\Phi},\tilde{\Phi},\tilde{\Phi})+\mathrm{higher\ orders},\end{equation}
where $\tilde{L}_1\equiv Q$ and $\tilde{L}_{n+1},n=0,1,2,...$ are composite closed string products with appropriately multiply NS and R states. We require that they are compatible with the $b_0^{-}$ and $L_0^{-}$ constraints,
\begin{equation}b_0^{-} \tilde{L}_{n+1} = 0,\ \ \ L_0^{-} \tilde{L}_{n+1} = 0,\end{equation}
satisfy $L_\infty$ relations, live in the small Hilbert space.

The composite products can be written as a sum of products of definite Ramond and picture number
\begin{equation}\tilde{L}_{n+1} = L_{n+1}^{(n)}|_0+L_{n+1}^{(n-1)}|_2+L_{n+1}^{(n-2)}|_4+...\ ,\end{equation}
where the sum terminates when the Ramond number exceeds the number of inputs. The Ramond number and picture number are correlated so that the NS part of the equations of motion has picture $-1$ and the R part of the equations of motion has picture $-1/2$. We want to build these products by inserting picture changing operators on a set of elementary products of odd degree at picture zero:
\begin{equation}L_1^{(0)}\equiv Q,\ \ \ L_2^{(0)},\ \ \ L_3^{(0)},\ \ \ L_4^{(0)},\ \ \ ...\ ,\end{equation}
which we assume are compatible with the $b_0^{-}$ and $L_0^{-}$ constraints,
\begin{equation}b_0^{-} L^{(0)}_{n+1} = 0,\ \ \ L_0^{-} L^{(0)}_{n+1} = 0,\end{equation}
live in the small Hilbert space, and satisfy $L_\infty$ relations. The most natural definition of the elementary products would derive from the polyhedral vertices of Saadi and Zwiebach \cite{Saadi}, but for our purposes it will not matter how they are chosen. 

The procedure for constructing the products is exactly like for the open string with stubs. We introduce a list of {\it products} and {\it gauge products}:
\begin{eqnarray}
\mathrm{products}:\lineup \ \ \ L_{d+p+r+1}^{(p)}|_{2r},\ \ \ \, \mathrm{degree\ odd},\nonumber\\
\mathrm{gauge\ products}:\lineup \ \ \ \lambda_{d+p+r+2}^{(p+1)}|_{2r},\ \ \ \ \ \mathrm{degree\ even},
\end{eqnarray} 
for $d,p,r\geq 0$. They are defined recursively following figure \ref{fig:Ramond2}, using the equations
\begin{eqnarray}
\lambda_{d+p+r+2}^{(p+1)}|_{2r} \lineup 
= \frac{d+1}{d+p+r+3}\Big(\xi_0 L_{d+p+r+2}^{(p)}|_{2r}-L_{d+p+r+2}^{(p)}|_{2r}(\xi_0\w \mathbb{I}_{d+p+r+1})\Big),\\
L_{d+p+r+2}^{(p+1)}|_{2r} \lineup = \frac{1}{p+1}\sum_{p'=0}^p\sum_{d'=0}^d\sum_{r'=0}^r \[ L_{d'+p'+r'+1}^{(p')}|_{2r'},\lambda_{d+p+r-d'-p'-r'+2}^{(p-p'+1)}|_{2(r-r')}\].
\end{eqnarray}
Note that the first equation uses the $\xi$ zero mode
\begin{equation}\xi_0 = \oint_{|z|=1}\frac{dz}{2\pi i}\frac{1}{z}\xi(z).\end{equation}
rather than an arbitrary operator built from $\xi$ as is possible for the open string. This guarantees that all products generated in the recursion are compatible with the $b_0^{-}$ and $L_0^{-}$ constraints. The proof that the resulting composite products $\tilde{L}_{n+1}$ are in the small Hilbert space and satisfy $L_\infty$ relations is identical to that of the previous section.

\section{Ramond Sectors of Type II Closed Superstring}

Now we discuss the equations of motion for type II closed superstring field theory. For type II closed superstrings, superconformal ghosts appear in both the holomorphic and antiholomorphic sectors. Therefore, string fields and closed string products will have two respective picture numbers, which we call left- or right-moving picture number. We need four dynamical closed string fields in the small Hilbert space, with respective left-/right-moving pictures:
\begin{eqnarray}
\Phi_{\mathrm{NN}}\lineup \in\mathcal{H}_{\mathrm{NN}},\ \ \ \mathrm{picture}\ \ (-1,-1),\\
\Psi_{\mathrm{NR}}\lineup\in\mathcal{H}_{\mathrm{NR}},\ \ \ \mathrm{picture}\ \ (-1,-1/2),\\
\Psi_{\mathrm{RN}}\lineup \in\mathcal{H}_{\mathrm{RN}},\ \ \ \mathrm{picture}\ \ (-1/2,-1),\\
\Phi_{\mathrm{RR}}\lineup\in\mathcal{H}_{\mathrm{RR}},\ \ \ \mathrm{picture}\ \ (-1/2,-1/2).
\end{eqnarray}
Here $\mathcal{H}_{\mathrm{NN}}$ is the NS-NS state space, $\mathcal{H}_{\mathrm{NR}}$ is the NS-R state space, $\mathcal{H}_{\mathrm{RN}}$ is the R-NS state space, and $\mathcal{H}_{\mathrm{RR}}$ is the R-R state space. All four string fields are degree even (and Grassmann even), ghost number 2, and satisfy the $b_0^{-}$ and level-matching constraints. The NS-NS and R-R string fields represent bosons, and are given as a linear combination of Grassmann even states with commuting coefficients. The NS-R and R-NS string fields represent fermions, and are given as a linear combination of Grassmann odd states with anti-commuting (Grassmann odd) coefficients. To be consistent with the Grassmann parity of the string field, we have to appropriately fix the Grassmannality of the Ramond ground states in type IIA and type IIB string theory. Using a bar to denote antiholomorphic or right-moving operators/quantities, we assume  
\begin{eqnarray} \lineup c \,\Theta_{\vec{s}}\,e^{-\phi/2}=\mathrm{Grassmann\ even},\nonumber\\
\lineup\overline{c} \,\overline{\Theta}_{\vec{s}}\,e^{-\overline{\phi}/2}=\mathrm{Grassmann\ even},
\end{eqnarray}
where in type IIB both holomorphic and antiholomorphic spinors have positive chirality, and in type IIA the holomorphic spinor has positive chirality and the antiholomorphic spinor has negative chirality. To construct the equations of motion we insert picture changing operators on a set of elementary products of odd degree at left- and right-moving picture zero:
\begin{equation}L_1^{(0,0)}\equiv Q,\ \ \ L_2^{(0,0)},\ \ \ L_3^{(0,0)},\ \ \ L_4^{(0,0)},\ \ \ ...\ .\end{equation}
We assume that these products satisfy $L_\infty$ relations, are compatible with the $b_0^{-}$ and $L_0^{-}$ constraints, and are in the small Hilbert space, meaning that they are separately annihilated by the left- and right-moving $\eta$ zero modes:
\begin{equation}[\eta,L_{n+1}^{(0,0)}]= 0,\ \ \ \ [\overline{\eta},L_{n+1}^{(0,0)}]= 0.\end{equation}
As in the heterotic string, we can take these products to be defined by the polyhedral closed string vertices of Saadi and Zwiebach. The construction of the equations of motion requires the definition of many products at intermediate Ramond and picture numbers. We will indicate the number of inputs, the left-/right-moving picture number, and the left-/right-moving Ramond number of products through indices as follows:
\begin{equation}
\setlength{\unitlength}{.25cm}
\begin{picture}(18,12)
\put(1.75,10.25){\vector(0,-3){2.5}}
\put(3.75,10.25){\oval(4,2)[tl]}
\put(4,11){left-moving picture number}
\put(2.5,8.5){\vector(0,-3){.75}}
\put(4.5,8.5){\oval(4,2)[tl]}
\put(4.75,9.25){right-moving picture number}
\put(0,6){$L_{N+1}^{(p,\overline{p})}|_{2r,2\overline{r}}$.}
\put(2.5,1.25){\vector(0,1){4}}
\put(4.5,1.25){\oval(4,2)[bl]}
\put(4.75,0){number of inputs}
\put(4.5,3){\vector(0,1){2.25}}
\put(6.5,3){\oval(4,2)[bl]}
\put(6.75,1.75){left-moving Ramond number}
\put(5.75,4.75){\vector(0,1){.5}}
\put(7.75,4.75){\oval(4,2)[bl]}
\put(8,3.5){right-moving Ramond number}
\end{picture}
\end{equation}
Left-moving Ramond number is defined as the number of holomorphic Ramond inputs minus the number of holomorphic Ramond outputs, and right-moving Ramond number is defined as the number of anti-holomorphic Ramond inputs minus anti-holomorphic Ramond outputs. 

Let us write the equations of motion explicitly out to second order:
\begin{eqnarray} 
\lineup  
0 = Q\Phi_{\mathrm{NN}}+L_2^{(1,1)}(\Phi_{\mathrm{NN}},\Phi_{\mathrm{NN}}) + L_2^{(1,0)}(\Psi_{\mathrm{NR}},\Psi_{\mathrm{NR}}) \nonumber\\
\lineup \ \ \ \ \ \ \ \ \ \ \ \ \ \  \,
+ L_2^{(0,1)}(\Psi_{\mathrm{RN}},\Psi_{\mathrm{RN}})+L_2^{(0,0)}(\Phi_{\mathrm{RR}},\Phi_{\mathrm{RR}})+\mathrm{higher\ orders}, \\
\lineup
0 = Q\Psi_{\mathrm{NR}} + 2L_2^{(1,1)}(\Phi_{\mathrm{NN}},\Psi_{\mathrm{NR}})+2L_2^{(0,1)}(\Psi_{\mathrm{RN}},\Phi_{\mathrm{RR}})+\mathrm{higher\ orders},\\
\lineup
0 = Q\Psi_{\mathrm{RN}} + 2L_2^{(1,1)}(\Phi_{\mathrm{NN}},\Psi_{\mathrm{RN}})+2L_2^{(1,0)}(\Psi_{\mathrm{NR}},\Phi_{\mathrm{RR}})+\mathrm{higher\ orders},\\
\lineup
0 = Q\Phi_{\mathrm{RR}} + 2L_2^{(1,1)}(\Phi_{\mathrm{NN}},\Phi_{\mathrm{RR}})+2L_2^{(1,1)}(\Psi_{\mathrm{RN}},\Psi_{\mathrm{NR}})+\mathrm{higher\ orders}.
\end{eqnarray}
The factors of $2$ in the last three equations arise from the fact that the fields can appear in either order in the 2-product. Note that the picture numbers of the products are fixed consistently with the picture of the linear terms. The products $L_2^{(1,1)},L_2^{(1,0)}$ and $L_2^{(0,1)}$ can be taken from \cite{ClosedSS} and are defined:
\begin{eqnarray}
L_2^{(1,1)}(A_1,A_2)\lineup = \frac{1}{9}\Big(X_0\overline{X}_0L_2^{(0,0)}(A_1,A_2) + X_0L_2^{(0,0)}(\overline{X}_0A_1,A_2)+X_0
L_2^{(0,0)}(A_1,\overline{X}_0A_2)\nonumber\\
\lineup\ \ \ \ \ \ \ \ \overline{X}_0L_2^{(0,0)}(X_0A_1,A_2) + L_2^{(0,0)}(X_0\overline{X}_0A_1,A_2)+
L_2^{(0,0)}(X_0 A_1,\overline{X}_0A_2)\nonumber\\
\lineup\ \ \ \ \ \ \ \ \overline{X}_0L_2^{(0,0)}(A_1,X_0A_2) + X_0L_2^{(0,0)}(\overline{X}_0A_1,X_0A_2)+
L_2^{(0,0)}( A_1,X_0\overline{X}_0A_2)\Big),\\
L_2^{(1,0)}(A_1,A_2)\lineup = \frac{1}{3}\Big(X_0L_2^{(0,0)}(A_1,A_2) + L_2^{(0,0)}(X_0A_1,A_2)+
L_2^{(0,0)}(A_1,X_0A_2)\Big),\\
L_2^{(0,1)}(A_1,A_2)\lineup = \frac{1}{3}\Big(\overline{X}_0L_2^{(0,0)}(A_1,A_2) + L_2^{(0,0)}(\overline{X}_0A_1,A_2)+
L_2^{(0,0)}(A_1,\overline{X}_0A_2)\Big),
\end{eqnarray}
where
\begin{equation}
X_0 = [Q,\xi_0],\ \ \ \ \overline{X}_0 = [Q,\overline{\xi}_0],
\end{equation}
and $\xi_0$ and $\overline{\xi}_0$ are the holomorphic and antiholomorphic $\xi$ zero modes:
\begin{equation}\xi_0=\oint_{|z|=1}\frac{dz}{2\pi i}\frac{\xi(z)}{z},\ \ \ \ \overline{\xi}_0=\oint_{|\overline{z}|=1}\frac{d\overline{z}}{2\pi i}\frac{\overline{\xi}(\overline{z})}{\overline{z}}.\end{equation}
These products are the natural closed superstring counterparts of \eq{M2stub}.

\begin{figure}
\begin{center}
\resizebox{3in}{1.5in}{\includegraphics{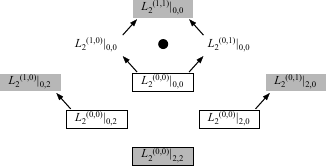}}
\end{center}
\caption{\label{fig:Ramond4}  Diagram showing the construction of the 2-products of type II closed superstring field theory. Products shaded in grey appear in the equations of motion, and the elementary products at picture zero are boxed. Arrows pointing to the left indicate the construction of a left gauge product out of the product at the source of the arrow, which then feeds into the definition of the product at the target of the arrow. Arrows pointing to the right indicate the construction of a right gauge product out of the product at the source of the arrow, which then feeds into the definition of the product at the target of the arrow.}
\end{figure}

Let us introduce the composite string field 
\begin{equation}\tilde{\Phi} = \Phi_{\mathrm{NN}}+\Psi_{\mathrm{NR}}+\Psi_{\mathrm{RN}}+\Phi_{\mathrm{RR}}\in\tilde{\mathcal{H}}.\end{equation}
We can write the equations of motion in the form
\begin{equation}
0=Q\tilde{\Phi}+\tilde{L}_2(\tilde{\Phi},\tilde{\Phi})+\tilde{L}_3(\tilde{\Phi},\tilde{\Phi},\tilde{\Phi})+\mathrm{higher\ orders},
\end{equation}
where $\tilde{L}_{n+1}$ are composite products of odd degree which appropriately multiply the four sectors of states. We require that the composite products satisfy $L_\infty$ relations and are in the small Hilbert space. The composite products can be decomposed into a sum of products with definite left- and right-moving Ramond and picture numbers:
\begin{eqnarray}
\tilde{L}_{n+1} \lineup = \ \ L_{n+1}^{(n,n)}|_{0,0}\  \ \ + L_{n+1}^{(n,n-1)}|_{0,2}\ \ \ + L_{n+1}^{(n,n-2)}|_{0,4}+...\nonumber\\
\lineup \ \ \  +L_{n+1}^{(n-1,n)}|_{2,0}+ L_{n+1}^{(n-1,n-1)}|_{2,2}+...\nonumber\\
\lineup\ \ \ + L_{n+1}^{(n-2,n)}|_{4,0}+...\nonumber\\
\lineup\ \ \ +...\ ,
\end{eqnarray}
where the double sum terminates when the left- or right-moving Ramond number exceeds the number of inputs. The left/right Ramond and picture numbers of the products are correlated so that the NS-NS part of the equations of motion has picture $(-1,-1)$, the NS-R part of the equations of motion has picture $(-1,-1/2)$, the R-NS part of the equations of motion has picture $(-1/2,-1)$, and the R-R part of the equations of motion has picture $(-1/2,-1/2)$. For example, at the first few orders the composite products take the form
\begin{eqnarray}
\tilde{L}_1 \lineup \equiv Q,\nonumber\\
\tilde{L}_2\lineup = \ \ L_2^{(1,1)}|_{0,0}+L_2^{(1,0)}|_{0,2}\nonumber\\
\lineup\ \ \ +L_2^{(0,1)}|_{2,0}+L_2^{(0,0)}|_{2,2},\nonumber\\
\tilde{L}_3\lineup = \ \ L_3^{(2,2)}|_{0,0}+L_3^{(2,1)}|_{0,2}\nonumber\\
\lineup\ \ \ +L_3^{(1,2)}|_{2,0}+L_3^{(1,1)}|_{2,2},\nonumber\\
\tilde{L}_4\lineup = \ \ L_4^{(3,3)}|_{0,0}+L_4^{(3,2)}|_{0,2}+L_4^{(3,1)}|_{0,4}\nonumber\\
\lineup\ \ \ +L_4^{(2,3)}|_{2,0}+L_4^{(2,2)}|_{2,2}+L_4^{(2,1)}|_{2,4}\nonumber\\
\lineup\ \  \ +L_4^{(1,3)}|_{4,0}+L_4^{(1,2)}|_{4,2}+L_4^{(1,1)}|_{4,4}.\nonumber\\
\end{eqnarray}
The products of $\tilde{L}_2$ were described in the previous paragraph. Nevertheless, let us explain how to ``construct" them in a way that generalizes to higher orders. The product $L_2^{(0,0)}|_{2,2}$ does not need to be constructed; it is already provided by the elementary 2-product acting on two R-R states:
\begin{equation}L_2^{(0,0)}|_{2,2} = \mathrm{given}.\end{equation}
The products $L_2^{(1,0)}|_{0,2}$ and $L_2^{(1,0)}|_{0,2}$ can be derived by climbing ``ladders"
\begin{eqnarray}
L_2^{(0,0)}|_{0,2}\lineup =\mathrm{given},\\
\lambda_2^{(1,0)}|_{0,2}\lineup = \frac{1}{3}\Big(\xi_0L_2^{(0,0)}|_{0,2} +L_2^{(0,0)}|_{0,2}(\xi_0\w\mathbb{I})\Big),\\
L_2^{(1,0)}|_{0,2}\lineup = \[Q,\lambda_2^{(1,0)}|_{0,2}\],
\end{eqnarray}
and 
\begin{eqnarray}
L_2^{(0,0)}|_{2,0}\lineup =\mathrm{given},\\
\overline{\lambda}_2^{(0,1)}|_{2,0}\lineup = \frac{1}{3}\Big(\overline{\xi}_0L_2^{(0,0)}|_{2,0} +L_2^{(0,0)}|_{2,0}(\overline{\xi}_0\w\mathbb{I})\Big),\\
L_2^{(0,1)}|_{2,0}\lineup =\[Q,\overline{\lambda}_2^{(0,1)}|_{2,0}\].
\end{eqnarray}
The products denoted with $\lambda$ and $\overline{\lambda}$ will be called {\it left} and {\it right} gauge products, respectively. Following \cite{ClosedSS}, $L_2^{(1,1)}|_{0,0}$ can be constructed as follows:\footnote{Our approach generalizes the symmetric construction of the NS-NS sector described in \cite{ClosedSS}. The generalization of the asymmetric construction appears to be less direct and we did not consider it. } 
\begin{eqnarray}
\lineup\ \ \ \ \ \ \ \ \ \  \ \ \ \ \ \ \ \ \ \ \ \ \ \ \ \ \ \ \  \ \ \ \ \ \ \ \ \ 
 L_2^{(0,0)}|_{0,0}=\mathrm{given},\\
\lineup\ \ \ \ \ \ \ \ \ \  \ \ \ \ \ \ \ \ \ \ \ \ \ \ \ \ \ \ \  \ \ \ \ \ \ \ \ \ 
\setlength{\unitlength}{.25cm}
\begin{picture}(10,0)
\put(0,1){\vector(-1,-1){1.75}}
\put(10,1){\vector(1,-1){1.75}}
\end{picture}\nonumber\\
\lambda_2^{(1,0)}|_{0,0} \lineup = \frac{1}{3}\Big(\xi_0L_2^{(0,0)}|_{0,0} +L_2^{(0,0)}|_{0,0}(\xi_0\w\mathbb{I})\Big),\ \ \ \ \ \ \ \ \ 
\overline{\lambda}_2^{(0,1)}|_{0,0} = \frac{1}{3}\Big(\overline{\xi}_0L_2^{(0,0)}|_{0,0} +L_2^{(0,0)}|_{0,0}(\overline{\xi}_0\w\mathbb{I})\Big),\\
L_2^{(1,0)}|_{0,0}\lineup = \[Q,\lambda_2^{(1,0)}|_{0,0}\],\ \ \ \ \ \ \ \ \ \ \ \ \ \ \ \ \ \ \ \ \ \ \ \ \ \ \ \ \ \ \ \ \ \ \ \ 
L_2^{(0,1)}|_{0,0} =\[Q,\overline{\lambda}_2^{(0,1)}|_{0,0}\],\\
\lambda_2^{(1,1)}|_{0,0} \lineup = \frac{1}{3}\Big(\xi_0 L_2^{(1,0)}|_{0,0} +L_2^{(1,0)}|_{0,0}(\xi_0\w\mathbb{I})\Big),\ \ \ \ \ \ \ \ \ 
\overline{\lambda}_2^{(1,1)}|_{0,0} = \frac{1}{3}\Big(\overline{\xi}_0L_2^{(0,1)}|_{0,0} +L_2^{(0,1)}|_{0,0}(\overline{\xi}_0\w\mathbb{I})\Big),\\
\lineup\ \ \ \ \ \ \ \ \ \  \ \ \ \ \ \ \ \ \ \ \ \ \ \ \ \ \ \ \  \ \ \ \ \ \ \ \ \ 
\setlength{\unitlength}{.25cm}
\begin{picture}(10,0)
\put(-2,1){\vector(1,-1){1.75}}
\put(11,1){\vector(-1,-1){1.75}}
\end{picture}\nonumber\\
\lineup\ \ \ \ \ \ \ \ \ \  \ \ \ \ \ \ \ \ \ \ \ \ \ \ \ \ \ \ \  
 L_2^{(1,1)}|_{0,0} = \frac{1}{2}\[Q,\lambda_2^{(1,1)}|_{0,0}+\overline{\lambda}_2^{(1,1)}|_{0,0}\].
\end{eqnarray}
Note that the procedure splits into two independent threads. This is the first indication that the recursion defining the products is effectively 2-dimensional, reflecting the need to insert picture in both the left- and right-moving sectors. We illustrate the construction of 2-string products diagrammatically in figure \ref{fig:Ramond4}. In particular, the above construction of $L_2^{(1,1)}|_{0,0}$ corresponds to filling a ``diamond" of products of intermediate left- and right-moving picture number.

\begin{figure}
\begin{center}
\resizebox{4in}{2in}{\includegraphics{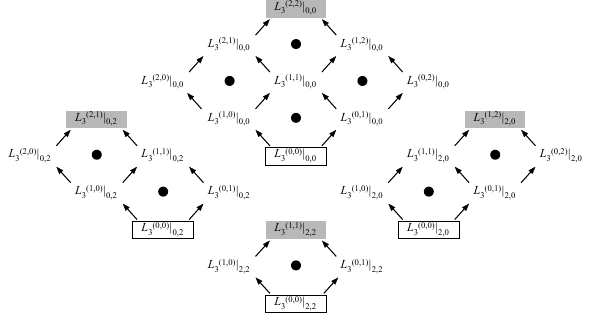}}
\end{center}
\caption{\label{fig:Ramond5}  Diagram showing the construction of the 3-products of type II closed superstring field theory.}
\end{figure}

\begin{figure}
\begin{center}
\resizebox{6.5in}{3.2in}{\includegraphics{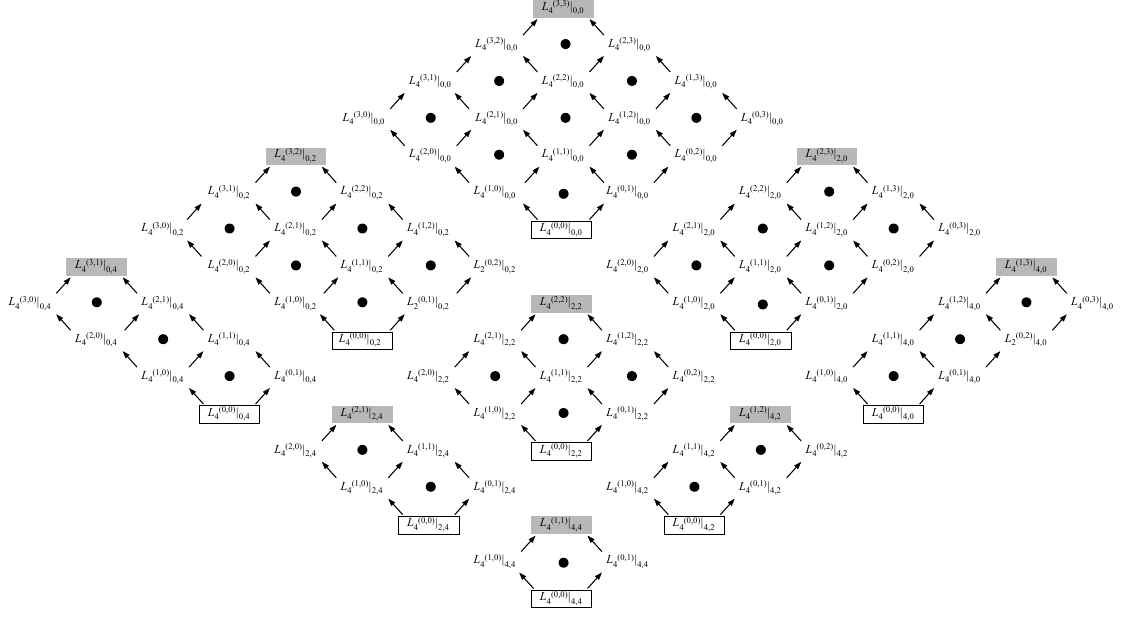}}
\end{center}
\caption{\label{fig:Ramond6}  Diagram showing the construction of the 4-products of type II closed superstring field theory.}
\end{figure}

Proceeding to the next order, the construction of 3-string products is illustrated in figure \ref{fig:Ramond5}. This requires filling two ``diamonds" to construct $L_3^{(2,2)}|_{0,0}$ and $L_3^{(1,1)}|_{2,2}$, and two ``prisms" to construct $L_3^{(2,1)}|_{0,2}$ and $L_3^{(1,2)}|_{2,0}$. To emphasize the pattern, we also illustrate the construction of 4-string products in figure \ref{fig:Ramond6}. The explicit formulas that go along with these diagrams are easiest to motivate by proceeding directly to the differential equations for the generating functions. We start with a list of all required products and left/right gauge products:
\begin{eqnarray}
\mathrm{products:}\lineup \ \ \ L_{N+1}^{(p,\overline{p})}|_{2r,2\overline{r}},\nonumber\\
\mathrm{left\ gauge\ products}:\lineup \ \ \ \lambda_{N+2}^{(p+1,\overline{p})}|_{2r,2\overline{r}},\nonumber\\
\mathrm{right\ gauge\ products}:\lineup \ \ \ \overline{\lambda}_{N+2}^{(p,\overline{p}+1)}|_{2r,2\overline{r}},
\end{eqnarray}
The integers $N,p,r,\overline{p},\overline{r}$ take the respective ranges
\begin{eqnarray}
 L_{N+1}^{(p,\overline{p})}|_{2r,2\overline{r}}:\lineup \ \ \ N\geq 0;\ \ \  \begin{matrix}0\leq r\leq N,\ \ \ \ \ \ \ \ 0\leq p\leq N-r \\
0\leq \overline{r}\leq N,\ \ \ \ \ \ \ \ 0\leq \overline{p}\leq N-\overline{r}\end{matrix},\nonumber\\
\lambda_{N+2}^{(p+1,\overline{p})}|_{2r,2\overline{r}}:\lineup \ \ \ N\geq 0;\ \ \  \begin{matrix}0\leq r\leq N,\ \ \ \ \ \ \ \ 0\leq p\leq N-r \ \ \ \ \ \,\\
0\leq \overline{r}\leq N+1,\ \ \ 0\leq \overline{p}\leq N-\overline{r}+1\end{matrix},\nonumber\\
\overline{\lambda}_{N+2}^{(p,\overline{p}+1)}|_{2r,2\overline{r}}:\lineup \ \ \ N\geq 0;\ \ \  \begin{matrix}0\leq r\leq N+1,\ \ \ 0\leq p\leq N-r+1 \\
0\leq \overline{r}\leq N,\ \ \ \ \ \ \ \  0\leq \overline{p}\leq N-\overline{r}\ \ \ \ \ \end{matrix}.
\label{eq:TypeIIlist}\end{eqnarray}
The allowed range of $r$ and $\overline{r}$ is generous, as some products and left/right gauge products listed above vanish because the left- or right-moving Ramond number exceeds the number of inputs. Ignoring products which vanish, one can double check that this list maps onto the diagrams of figures \ref{fig:Ramond4}-\ref{fig:Ramond6} (note that left and right arrows represent left and right gauge products, respectively). In defining the generating functions, at first one might guess that we need six parameters $s,\overline{s},t,\overline{t}$ and $u,\overline{u}$, counting left/right picture deficit, left/right picture, and left/right Ramond number, respectively. This is almost correct, but would be over-counting since the products only have five integer labels. This is a consequence of the fact that the left/right picture deficits, picture and Ramond numbers are not all independent: they are constrained by the fact that the rank of multiplication in the holomorphic and antiholomorphic sectors is the same. The correct way to deal with this is to identify $t=\overline{t}$, so the generating functions depend on only five parameters $s,\overline{s},t,u,\overline{u}$. Promoting the products and left/right gauge products to coderivations on the symmetrized tensor algebra, we therefore define generating functions
\begin{eqnarray}
\L(s,\overline{s},t,u,\overline{u})\lineup = \sum_{N=0}^{\infty}\,\sum_{d,p,r=0}^{\infty}\,\sum_{\overline{d},\overline{p},\overline{r}=0}^{\infty} (s^d\overline{s}^{\overline{d}})(t^{p+\overline{p}})(u^r\overline{u}^{\overline{r}})\delta_{d+p+r=N}\delta_{\overline{d}+\overline{p}+\overline{r}=N}\L_{N+1}^{(p,\overline{p})}|_{2r,2\overline{r}},\\
\ll(s,\overline{s},t,u,\overline{u})\lineup = \sum_{N=0}^{\infty}\,\sum_{d,p,r=0}^{\infty}\,\sum_{\overline{d},\overline{p},\overline{r}=0}^{\infty} (s^d\overline{s}^{\overline{d}})(t^{p+\overline{p}})(u^r\overline{u}^{\overline{r}})\delta_{d+p+r=N}\delta_{\overline{d}+\overline{p}+\overline{r}=N+1}\ll_{N+2}^{(p+1,\overline{p})}|_{2r,2\overline{r}}, \\
\llb(s,\overline{s},t,u,\overline{u})\lineup = \sum_{N=0}^{\infty}\,\sum_{d,p,r=0}^{\infty}\,\sum_{\overline{d},\overline{p},\overline{r}=0}^{\infty} (s^d\overline{s}^{\overline{d}})(t^{p+\overline{p}})(u^r\overline{u}^{\overline{r}})\delta_{d+p+r=N+1}\delta_{\overline{d}+\overline{p}+\overline{r}=N}\llb_{N+1}^{(p,\overline{p}+1)}|_{2r,2\overline{r}}.
\end{eqnarray}
We have inserted Kronecker deltas into the sums to enforce the appropriate relation between left/right picture deficit, picture and Ramond number and the number of inputs in the product.

Based on our results for the open string, it is natural to guess that the generating functions should satisfy the same set of differential equations which determine the products of the NS-NS sector \cite{ClosedSS} with extra parameters $u,\overline{u}$ that go along for the ride. In particular we should have
\begin{eqnarray}
\lineup\ \ \ \ \ \  \frac{\d}{\d t}\L(s,\overline{s},t,u,\overline{u}) = [\L(s,\overline{s},t,u,\overline{u}),\ll(s,\overline{s},t,u,\overline{u})+\llb(s,\overline{s},t,u,\overline{u})],\label{eq:TypeIIdiff1}\\
\lineup \ll(s,\overline{s},t,u,\overline{u})=\xi_0\circ\frac{\d}{\d s}\L(s,\overline{s},t,u,\overline{u}),\ \ \ \llb(s,\overline{s},t,u,\overline{u})=\overline{\xi}_0\circ\frac{\d}{\d \overline{s}}\L(s,\overline{s},t,u,\overline{u}).\label{eq:TypeIIdiff2}
\end{eqnarray}
Here the operation $\xi_0\circ$ is defined by its action on an $n$-closed string product
\begin{equation}\xi_0\circ b_n=\frac{1}{n+1}\Big(\xi_0 b_n+(-1)^{\deg(b_n)}b_n(\xi_0\w \mathbb{I}_{n-1})\Big).\end{equation}
This defines a homotopy operator for the coderivation $\n$:
\begin{equation}[\n,\xi_0\circ\b] +\xi_0\circ[\n,\b] = \b.\end{equation}
The same equations hold for $\overline{\xi}_0\circ$ with $\xi_0$ and $\n$ replaced with $\overline{\xi}_0$ and the coderivation for the right moving eta zero mode $\nb$. Next, we claim that at $t=0$ the generating function for products is nilpotent:
\begin{equation}[\L(s,\overline{s},0,u,\overline{u}),\L(s,\overline{s},0,u,\overline{u})]=0.\label{eq:Linf0}\end{equation}
Expanding in powers of $s,\overline{s},u,\overline{u}$, this equation is equivalent to
\begin{equation}
\sum_{n=0}^N \,\sum_{r'\in S_n}\,\sum_{\overline{r}'\in\overline{S}_n}[L_{n+1}^{(0,0)}|_{2r',2\overline{r}'},L_{N-n+1}^{(0,0)}|_{2(r-r'),2(\overline{r}-\overline{r}')}]=0,\label{eq:Linf0comp}
\end{equation}
where the sets $S_n$ and $\overline{S}_n$ defining the summations over $r'$ and $\overline{r}'$ are characterized by the conditions
\begin{eqnarray}
\lineup r'\in S_n\ \ \ \leftrightarrow\ \ \ 0\leq r',\ \ r-N+n\leq r',\ \  r'\leq n, \ \ r'\leq r,\\  
\lineup \overline{r}'\in \overline{S}_n\ \ \ \leftrightarrow\ \ \ 0\leq \overline{r}',\ \ \overline{r}-N+n\leq \overline{r}',\ \  \overline{r}'\leq n, \ \ \overline{r}'\leq \overline{r}.
\end{eqnarray}
To prove \eq{Linf0comp}, note that we can expand the range of summation on the left hand side 
\begin{equation}\sum_{n=0}^N \,\sum_{r'\in S_n}\,\sum_{\overline{r}'\in\overline{S}_n}[L_{n+1}^{(0,0)}|_{2r',2\overline{r}'},L_{N-n+1}^{(0,0)}|_{2(r-r'),2(\overline{r}-\overline{r}')}]=\sum_{n=0}^N \,\sum_{r'=0}^r\,\sum_{\overline{r}'=0}^{\overline{r}}[L_{n+1}^{(0,0)}|_{2r',2\overline{r}'},L_{N-n+1}^{(0,0)}|_{2(r-r'),2(\overline{r}-\overline{r}')}],
\end{equation}
since the additional terms all vanish on account of the fact that the left- or right-moving Ramond number of one of the products in the commutator exceeds the number of inputs. The sum over $r',\overline{r}'$ now reproduces the $2r,2\overline{r}$ projection of the $(N+1)$st $L_\infty$ relation for the elementary products, which gives zero:
\begin{eqnarray}
\sum_{n=0}^N \,\sum_{r'\in S_n}\,\sum_{\overline{r}'\in\overline{S}_n}[L_{n+1}^{(0,0)}|_{2r',2\overline{r}'},L_{N-n+1}^{(0,0)}|_{2(r-r'),2(\overline{r}-\overline{r}')}]\lineup =\sum_{n=0}^N [L_{n+1}^{(0,0)},L_{N-n+1}^{(0,0)}]|_{2r,2\overline{r}},\nonumber\\
\lineup = 0.
\end{eqnarray}
This proves \eq{Linf0}. Next note that 
\begin{equation}
\frac{\d}{\d t}[\L(s,\overline{s},t,u,\overline{u}),\L(s,\overline{s},t,u,\overline{u})] = [[\L(s,\overline{s},t,u,\overline{u}),\L(s,\overline{s},t,u,\overline{u})],\ll(s,\overline{s},t,u,\overline{u})+\llb(s,\overline{s},t,u,\overline{u})].
\end{equation}
Since this equation is homogeneous in $[\L(s,\overline{s},t,u,\overline{u}),\L(s,\overline{s},t,u,\overline{u})]$, which vanishes at $t=0$ as just demonstrated, we conclude
\begin{equation} [\L(s,\overline{s},t,u,\overline{u}),\L(s,\overline{s},t,u,\overline{u})] = 0.\end{equation}
Since the elementary products are in the small Hilbert space, we have
\begin{equation}[\n,\L(s,\overline{s},0,u,\overline{u})]=0,\ \ \ [\nb,\L(s,\overline{s},0,u,\overline{u})]=0.\end{equation}
Next consider 
\begin{eqnarray}
\frac{\d }{\d t}[\n,\L(s,\overline{s},t,u,\overline{u})] \lineup = [\n,[\L(s,\overline{s},t,u,\overline{u}),\ll(s,\overline{s},t,u,\overline{u})+\llb(s,\overline{s},t,u,\overline{u})]],\nonumber\\
\lineup = [[\n,\L(s,\overline{s},t,u,\overline{u})],\ll(s,\overline{s},t,u,\overline{u})+\ll(s,\overline{s},t,u,\overline{u})]
-[\L(s,\overline{s},t,u,\overline{u}),[\n,\ll(s,\overline{s},t,u,\overline{u})]]\nonumber\\
\lineup\ \ \ -[\L(s,\overline{s},t,u,\overline{u}),[\n,\llb(s,\overline{s},t,u,\overline{u})]]\nonumber,\\
\lineup = [[\n,\L(s,\overline{s},t,u,\overline{u})],\ll(s,\overline{s},t,u,\overline{u})+\ll(s,\overline{s},t,u,\overline{u})]
-\left[\L(s,\overline{s},t,u,\overline{u}),\frac{\d}{\d s}\L(s,\overline{s},t,u,\overline{u})\right]\nonumber\\
\lineup\ \ \ +\left[\L(s,\overline{s},t,u,\overline{u}),\xi_0\circ\frac{\d}{\d s}[\n,\L(s,\overline{s},t,u,\overline{u})]\right]+\left[\L(s,\overline{s},t,u,\overline{u}),\overline{\xi}_0\circ \frac{\d}{\d\overline{s}}[\n,\L(s,\overline{s},t,u,\overline{u})]\right],\nonumber\\
\lineup = [[\n,\L(s,\overline{s},t,u,\overline{u})],\ll(s,\overline{s},t,u,\overline{u})+\ll(s,\overline{s},t,u,\overline{u})]
-\frac{1}{2}\frac{\d}{\d s}[\L(s,\overline{s},t,u,\overline{u}),\L(s,\overline{s},t,u,\overline{u})]\nonumber\\
\lineup\ \ \ +\left[\L(s,\overline{s},t,u,\overline{u}),\xi_0\circ\frac{\d}{\d s}[\n,\L(s,\overline{s},t,u,\overline{u})]\right]+\left[\L(s,\overline{s},t,u,\overline{u}),\overline{\xi}_0\circ \frac{\d}{\d\overline{s}}[\n,\L(s,\overline{s},t,u,\overline{u})]\right],\nonumber\\
\lineup = [[\n,\L(s,\overline{s},t,u,\overline{u})],\ll(s,\overline{s},t,u,\overline{u})+\ll(s,\overline{s},t,u,\overline{u})]
+\left[\L(s,\overline{s},t,u,\overline{u}),\xi_0\circ\frac{\d}{\d s}[\n,\L(s,\overline{s},t,u,\overline{u})]\right]\nonumber\\
\lineup\ \ \ +\left[\L(s,\overline{s},t,u,\overline{u}),\overline{\xi}_0\circ \frac{\d}{\d\overline{s}}[\n,\L(s,\overline{s},t,u,\overline{u})]\right].\nonumber\\
\end{eqnarray}
Since this equation is homogeneous in $[\n,\L(s,\overline{s},t,u,\overline{u})]$, which vanishes at $t=0$, we conclude
\begin{equation}[\n,\L(s,\overline{s},t,u,\overline{u})] = 0,\end{equation}
and by a similar argument 
\begin{equation}[\nb,\L(s,\overline{s},t,u,\overline{u})] = 0.\end{equation}
Consider the coderivation representing the composite products which appear in the equations of motion:
\begin{equation}\tilde{\L} = \sum_{n=0}^{\infty}\tilde{\L}_{n+1} = \L(0,0,1,1,1).\end{equation}
The above results imply that 
\begin{equation}[\n,\tilde{\L}]=0,\ \ \ [\nb,\tilde{\L}]= 0,\ \ \ [\tilde{\L},\tilde{\L}] = 0,\end{equation}
which says that the composite products are in the small Hilbert space and satisfy $L_\infty$ relations. Finally, we can find the recursive equations for the products and left/right gauge products by expanding \eq{TypeIIdiff1} and \eq{TypeIIdiff2} in powers of $s,\overline{s},t,u,\overline{u}$:
\begin{eqnarray}
\lambda_{N+2}^{(p+1,\overline{p})}|_{2r,2\overline{r}} \lineup = \frac{N-p-r+1}{N+3}\Big(\xi_0 L_{N+2}^{(p,\overline{p})}|_{2r,2\overline{r}}-L_{N+2}^{(p,\overline{p})}|_{2r,2\overline{r}}(\xi_0\w\mathbb{I}_{N+1})\Big),\\
\lb_{N+2}^{(p,\overline{p}+1)}|_{2r,2\overline{r}} \lineup = \frac{N-\overline{p}-\overline{r}+1}{N+3}\Big(\overline{\xi}_0 L_{N+2}^{(p,\overline{p})}|_{2r,2\overline{r}}-L_{N+2}^{(p,\overline{p})}|_{2r,2\overline{r}}(\overline{\xi}_0\w\mathbb{I}_{N+1})\Big),\\
L_{N+2}^{(p,\overline{p})}|_{2r,2\overline{r}}\lineup = \frac{1}{p+\overline{p}}\left(\sum_{n=0}^N \,\sum_{(p',r')\in A_n}\,\sum_{(\overline{p}',\overline{r}')\in \overline{A}_n}[L_{n+1}^{(p',\overline{p}')}|_{2r',2\overline{r}'},\l_{N-n+2}^{(p-p',\overline{p}-\overline{p}')}|_{2(r-r'),2(\overline{r}-\overline{r}')}]\right.\nonumber\\
\lineup\ \ \ \ \ \ \ \ \ \ \ \ \ 
\left.+\sum_{n=0}^N \,\sum_{(p',r')\in B_n}\,\sum_{(\overline{p}',\overline{r}')\in \overline{B}_n}[L_{n+1}^{(p',\overline{p}')}|_{2r',2\overline{r}'},\lb_{N-n+2}^{(p-p',\overline{p}-\overline{p}')}|_{2(r-r'),2(\overline{r}-\overline{r}')}]\right).
\end{eqnarray}
The sums in the last equation are over all values of $p',r'$ and $\overline{p}',\overline{r}'$ such that the products and left/right gauge products in the commutators have admissible left- and right-moving Ramond and picture numbers, as defined by the list \eq{TypeIIlist}. Explicitly, $(p',r')$ is in a set $A_n$ or $B_n$ and $(\overline{p}',\overline{r}')$ is in a set $\overline{A}_n$ or $\overline{B}_n$ defined by the conditions:
\begin{eqnarray}
(p',r')\in A_n,\ \ \ \leftrightarrow\ \ \ \lineup 0\leq p',\ \ \ \  0\leq r',\ \ \ \ p'\leq p-1,\ \ \ \  r'\leq r,\nonumber\\
\lineup p+r-N+n-1\leq p'+r',\ \ \ \ p'+r'\leq N,\nonumber\\
(\overline{p}',\overline{r}')\in \overline{A}_n,\ \ \ \leftrightarrow\ \ \ \lineup 0\leq \overline{p}',\ \ \ \  0\leq \overline{r}',\ \ \ \ \overline{p}'\leq \overline{p},\ \ \ \  \overline{r}'\leq \overline{r},\nonumber\\
\lineup \overline{p}+\overline{r}-N+n\leq \overline{p}'+\overline{r}',\ \ \ \ \overline{p}'+\overline{r}'\leq N,\nonumber\\
(p',r')\in B_n,\ \ \ \leftrightarrow\ \ \ \lineup 0\leq p',\ \ \ \  0\leq r',\ \ \ \ p'\leq p,\ \ \ \  r'\leq r,\nonumber\\
\lineup p+r-N+n\leq p'+r',\ \ \ \ p'+r'\leq N,\nonumber\\
(\overline{p}',\overline{r}')\in \overline{B}_n,\ \ \ \leftrightarrow\ \ \ \lineup 0\leq \overline{p}',\ \ \ \  0\leq \overline{r}',\ \ \ \ \overline{p}'\leq \overline{p}-1,\ \ \ \  \overline{r}'\leq \overline{r},\nonumber\\
\lineup \overline{p}+\overline{r}-N+n-1\leq \overline{p}'+\overline{r}',\ \ \ \ \overline{p}'+\overline{r}'\leq N.
\end{eqnarray}
This completes the definition of the equations of motion of type II closed superstring field theory.

\section{Supersymmetry}
\label{sec:SUSY}

Now that we have equations of motion including all NS and R sectors, it is interesting to understand in what sense they are supersymmetric. Here we will limit our discussion to the open superstring using Witten's associative star product, leaving generalizations to future work. For heterotic and type II closed superstrings, spacetime supersymmetry can be described in a different way as part of the gauge symmetry at the interacting level \cite{1PIR}.  For other discussions of supersymmetry in open string field theory, see \cite{BerkRamond,WittenSSFT,democratic}.

We will work with open string field theory on a BPS D-brane with sixteen supersymmetries. As described in \cite{FMS}, the simplest way to realize a supersymmetry transformation is by integrating the zero-momentum fermion vertex in the $-1/2$ picture:
\begin{equation}s_1\equiv \sqrt{2}\oint_{|z|=1}\frac{dz}{2\pi i} \Theta_{\vec{s}}\,e^{-\phi/2}(z)\eps_{\vec{s}},\label{eq:s1}\end{equation}
where the index on $s_1$ indicates that this operator is a 1-string product. The object $\eps_{\vec{s}}$ is the supersymmetry parameter---a constant Grassmann odd spinor. To keep the notation simple, we leave the dependence of $s_1$ on the supersymmetry parameter implicit. We make a GSO($+$) projection in both NS and R sectors, so the supersymmetry parameter must have positive chirality. Then, following the Grassmann assignments in section \ref{sec:Witten}, the supersymmetry operator $s_1$ will be {\it Grassmann even} (and degree even).\footnote{The factor $\sqrt{2}$ is inserted to obtain the canonical normalization of the supersymmetry algebra.} It is natural to identify the linearized supersymmetry transformation of the NS field as
\begin{equation}\delta\Phi_\mathrm{N} = s_1 \Psi_\mathrm{R}.\label{eq:SN}\end{equation}
Likewise, the linearized supersymmetry transformation of the R field should be proportional to the NS field. However, to get the pictures to line up we need a supersymmetry operator of picture $+1/2$. This can be defined by integrating the zero momentum fermion vertex at picture $+1/2$:
\begin{equation}S_1 \equiv \sqrt{2}\oint_{|z|=1}\frac{dz}{2\pi i} \Big( i\d X_\mu \Theta_{\vec{s}'}e^{\phi/2}(z)\Gamma^{\mu}_{\vec{s}',\vec{s}} + b\eta\Theta_{\vec{s}}\,e^{3\phi/2}(z)\Big)\eps_{\vec{s}},\label{eq:S1}\end{equation}
where $\Gamma^\mu_{\vec{s},\vec{s}'}$ are 10 dimensional gamma matrices. This operator is also Grassmann even. The linearized supersymmetry transformation of the Ramond field will therefore be
\begin{equation}\delta \Psi_\mathrm{R} = S_1\Phi_{\mathrm{N}}.\label{eq:SR}\end{equation}
We can write both supersymmetry transformations in a single equation using the composite string field $\tilde{\Phi}=\Phi_\mathrm{N}+\Psi_\mathrm{R}$
\begin{equation}\delta\tilde{\Phi} = \tilde{S}_1\tilde{\Phi},\end{equation}
where
\begin{equation}\tilde{S}_1 = S_1|_{-1} + s_1|_1,\label{eq:tS1}\end{equation}
with the indicated Ramond projections of $S_1$ and $s_1$. Now the question is how this supersymmetry generalizes to the nonlinear theory.

Before we get to this, however, let us mention a few essential properties of $s_1$ and $S_1$. Both operators are in the small Hilbert space and BRST invariant
\begin{eqnarray}
[\eta,s_1] \lineup = [\eta,S_1] = 0,\\
\ [Q,s_1]\lineup = [Q,S_1] = 0.
\end{eqnarray}
This is in fact necessary for the supersymmetry transformation to be a symmetry of the linearized equations of motion. Moreover, since both operators are zero modes of weight one primary fields, they are derivations of Witten's open string star product:
\begin{equation}[s_1,m_2] = [S_1,m_2] = 0.\end{equation}
It will also be useful to introduce a Grassmann odd supersymmetry operator in the large Hilbert space:
\begin{equation}
\sigma_1 \equiv \sqrt{2}\oint_{|z|=1}\frac{dz}{2\pi i} \xi \Theta_{\vec{s}}\,e^{-\phi/2}(z)\eps_{\vec{s}}.
\end{equation}
This operator satisfies
\begin{eqnarray}
S_1\lineup = [Q,\sigma_1],\\
s_1\lineup = [\eta,\sigma_1].
\end{eqnarray}
and is also a derivation of the star product. Note the relationship between the supersymmetry operators $S_1,\sigma_1$ and $s_1$ is somewhat analogous to the relation between the products $M_2,\mu_2$, and $m_2$.

\subsection{Perturbative Construction of Supersymmetry Transformation}

In this subsection we give a perturbative construction of the supersymmetry transformation. It turns out that the final form of the supersymmetry transformation is easiest to understand in a different set of field variables, described in the next subsection, and some readers may wish to skip ahead. (See, in particular, equation \eq{q}) The following derivation, however, has the advantage that it likely generalizes to supersymmetry transformations in other forms of superstring field theory. 

Our task is to find a sequence of degree even products in the small Hilbert space,
\begin{equation}\tilde{S}_1,\ \ \ \tilde{S}_2,\ \ \ \tilde{S}_3,\ \ \ ...\ ,\end{equation}
so that the supersymmetry transformation can be written
\begin{equation}\delta \tilde{\Phi} = \tilde{S}_1\tilde{\Phi} + \tilde{S}_2(\tilde{\Phi},\tilde{\Phi}) + \tilde{S}_3(\tilde{\Phi},\tilde{\Phi},\tilde{\Phi})+\mathrm{higher\ orders}.\label{eq:tSall}\end{equation}
Let us explain what it means for this transformation to be a symmetry. Consider a pair of coderivations $\tilde{\M}$ and $\tS$ representing the products in the equations of motion and supersymmetry transformation: 
\begin{eqnarray}\tilde{\M} \lineup \equiv \sum_{n=0}^{\infty} \tilde{\M}_{n+1},\\
\tS \lineup \equiv \sum_{n=0}^{\infty} \tS_{n+1},
\end{eqnarray}
where $\tilde{M}_{n+1}$ and related products in the equations of motion are taken from section \ref{sec:Witten}. We also introduce a group-like element\footnote{See \cite{WB} for more detailed discussion of coderivations, cohomomorphisms, and group-like elements.}
\begin{equation}\frac{1}{1-\tilde{\Phi}} = \, 1_{T\tilde{\mathcal{H}}}\,+\, \tilde{\Phi}\, +\, \tilde{\Phi}\otimes\tilde{\Phi}\, + \, \tilde{\Phi}\otimes\tilde{\Phi}\otimes\tilde{\Phi}\, +\, ...\ .\end{equation}
The equations of motion and the supersymmetry transformation can be expressed in the form:
\begin{equation}
\tilde{\M}\frac{1}{1-\tilde{\Phi}} = 0, \ \ \ \ \ \ 
\delta \frac{1}{1-\tilde{\Phi}} = \tS\frac{1}{1-\tilde{\Phi}}.
\end{equation}
We claim that the products $\tilde{S}_{n+1}$ generate a symmetry if  
\begin{equation}[\tilde{\M},\tS] = 0.\label{eq:SUSY}\end{equation}
To see this, suppose $\tilde{\Phi}$ is a solution to the equations of motion. Then calculate the equations of motion on the transformed string field $\tilde{\Phi}+\delta\tilde{\Phi}$:    
\begin{equation}\tilde{\M}\left(\frac{1}{1-\tilde{\Phi}}+\delta\frac{1}{1-\tilde{\Phi}}\right) = \tilde{\M}\tS\frac{1}{1-\tilde{\Phi}} = \tS\tilde{\M}\frac{1}{1-\tilde{\Phi}} = 0.\end{equation}
Therefore the transformation is a symmetry. Note that the condition $[\tilde{\M},\tS]=0$ is somewhat stronger than the statement that the transformation maps solutions into solutions. It places a nontrivial condition on the off-shell form of the supersymmetry transformation. In fact, \eq{SUSY} is the nearest we can come to the statement that the transformation is a symmetry of the action---all that is missing is a symplectic structure which would allow us to define an action and impose cyclicity. 

Equation \eq{SUSY} implies that the products $\tilde{S}_{n+1}$ satisfy a hierarchy of identities:
\begin{equation}[\tilde{M}_1,\tilde{S}_{n+1}]+[\tilde{M}_{2},\tilde{S}_n]+... + [\tilde{M}_{n+1},\tilde{S}_1]  =0,\ \ \ \ n=0,1,2,...\ ,\end{equation}
where $\tilde{M}_1=Q$. We have already discussed $\tilde{S}_1$ which appears in \eq{tS1}, and it provides a solution to
\begin{equation}[Q,\tilde{S}_1]=0.\end{equation}
The next step is to derive the 2-product $\tilde{S}_2$ from the identity
\begin{equation}
[Q,\tilde{S}_2] + [\tilde{M}_2,\tilde{S}_1]=0.\label{eq:2ndid}
\end{equation}
Since $\tilde{S}_2$ is part of a supersymmetry transformation, it can be split into a sum of products with odd Ramond number:
\begin{equation}\tilde{S}_2 = S_2|_{-1}+ s_2|_1.\end{equation}
Equation \eq{2ndid} breaks up into two independent equations at Ramond number $-1$ and $1$:
\begin{eqnarray}
0\lineup =\[Q,S_2|_{-1}] + \[M_2|_0,S_1|_{-1}\],\\
0\lineup = \[Q,s_2|_1\] + \[M_2|_0,s_1|_1\] + \[m_2|_2,S_1|_{-1}\].\label{eq:s21rel}
\end{eqnarray}
To solve the first equation, we can pull a $Q$ out of either $M_2$ or $S_1$. The solution we prefer is to pull a $Q$ out of $M_2$, obtaining 
\begin{equation}S_2|_{-1} = \[S_1|_{-1},\mu_2|_0\].\end{equation}
We can check that this is in the small Hilbert space:
\begin{eqnarray}
\[\eta,S_2|_{-1}\] = \[S_1|_{-1},m_2|_0\]=[S_1,m_2]|_{-1}=0,
\end{eqnarray}
where we used conservation of Ramond number and the fact that $S_1$ is a derivation of $m_2$. To solve for the Ramond number $1$ component of $\tilde{S}_2$, we pull a $Q$ in the natural way out of the two terms in \eq{s21rel}
\begin{equation}s_2|_1 = \[s_1|_1,\mu_2|_0\] + \[m_2|_2,\sigma_1|_{-1}\].\label{eq:s21}\end{equation}
It turns out to be convenient to introduce separate symbols for these two terms: 
\begin{eqnarray} s_2^{(\mathrm{I})}|_1 \lineup \equiv \[s_1|_1,\mu_2|_0\],\\
s_2^{(\mathrm{II})}|_1\lineup \equiv \[m_2|_2,\sigma_1|_{-1}\].\end{eqnarray}
Check that this is in the small Hilbert space:
\begin{eqnarray}\[\eta,s_2|_1\] \lineup = \[s_1|_1,m_2|_0\] - \[m_2|_2,s_1|_{-1}\] ,\nonumber\\
\lineup = [s_1,m_2]|_1,\nonumber\\
\lineup = 0,\end{eqnarray}
where we used conservation of Ramond number and the fact that $s_1$ is a derivation of $m_2$. This completes the definition of the 2-product $\tilde{S}_2$ in the supersymmetry transformation.

To establish a pattern it is helpful to continue on to the 3-product $\tilde{S}_3$, which must satisfy
\begin{equation}[Q,\tilde{S}_3]+[\tilde{M}_2,\tilde{S}_2]+[\tilde{M}_3,\tilde{S}_1]=0.\label{eq:tS3rel}\end{equation}
Let us look at the component of this equation at Ramond number 3:
\begin{equation}\[Q,\tilde{S}_3|_3\]+\[m_2|_2,s_2|_1\]+\[m_3'|_2,s_1|_1\]=0.\end{equation}
Substituting \eq{s21} for $s_2|_1$ and \eq{m3p2} for $m_3'|_2$,
\begin{eqnarray}
0\lineup =\[Q,\tilde{S}_3|_3\]+\[m_2|_2,\[s_1|_1,\mu_2|_0\]\]+\[m_2|_2,\[m_2|_2,\sigma_1|_{-1}\]\]+\[\[m_2|_2,\mu_2|_0\],s_1|_1\],\nonumber\\
\lineup = \[Q,\tilde{S}_3|_3\],
\end{eqnarray}
where the additional terms either cancel or vanish identically because the Ramond number exceeds the number of inputs. From this we conclude that the Ramond number 3 component of $\tilde{S}_3$ can be set to zero. Therefore $\tilde{S}_3$ must be a sum of products at Ramond number $-1$ and $1$ 
\begin{equation}\tilde{S}_3 = S_3|_{-1} + s_3|_1,\end{equation}
just like $\tilde{S}_1$ and $\tilde{S}_2$. Let's look at the Ramond number $-1$ component of the identity \eq{tS3rel}:
\begin{eqnarray}0\lineup =\[Q,S_3|_{-1}\]+\[M_2|_0,S_2|_{-1}\]+\[M_3|_0,S_1|_{-1}\],\nonumber\\
\lineup = \[Q,S_3|_{-1}\]+\[M_2|_0,S_2|_{-1}\]+\frac{1}{2}\[\[Q,\mu_3|_0\]+\[M_2|_0,\mu_2|_0\],S_1|_{-1}\],\nonumber\\
\lineup = \[Q,S_3|_{-1}\]+\frac{1}{2}\[Q,\[\mu_3|_0,S_1|_{-1}\]\]+\[M_2|_0,S_2|_{-1}\]-\frac{1}{2}\[M_2|_0,S_2|_{-1}\]+\frac{1}{2}\[\[M_2|_0,S_1|_{-1}\],\mu_2|_0\],\nonumber\\
\lineup = \[Q,S_3|_{-1}\]+\frac{1}{2}\[Q,\[\mu_3|_0,S_1|_{-1}\]\]+\frac{1}{2}\[M_2|_0,S_2|_{-1}\]+\frac{1}{2}\[\[Q,S_2|_{-1}],\mu_2|_0\],\nonumber\\
\lineup = \left[Q,S_3|_{-1} - \frac{1}{2}\Big(\[S_1|_{-1},\mu_3|_0\]+\[S_2|_{-1},\mu_2|_0\]\Big)\right].
\end{eqnarray}
This suggests we identify 
\begin{equation}S_3|_{-1} = \frac{1}{2}\Big(\[S_1|_{-1},\mu_3|_0\]+\[S_2|_{-1},\mu_2|_0\]\Big).\end{equation}
Check that this is in the small Hilbert space:
\begin{eqnarray}
\[\eta,S_3|_{-1}\] \lineup = \frac{1}{2}\Big(\[S_1|_{-1},m_3|_0\]+\[S_2|_{-1},m_2|_0\]\Big),\nonumber\\
\lineup = \frac{1}{2}\Big(\[S_1|_{-1},\[m_2|_0,\mu_2|_0\]\]+\[S_2|_{-1},m_2|_0\]\Big),\nonumber\\
\lineup = \frac{1}{2}\Big(\[m_2|_0,S_2|_{-1}\]+\[S_2|_{-1},m_2|_0\]\Big),\nonumber\\
\lineup = 0.
\end{eqnarray}
Finally, let's look at the Ramond number 1 component of \eq{tS3rel}:
\begin{eqnarray}0\lineup =\[Q,s_3|_{1}\]+\[M_2|_0,s_2|_1\]+\[m_2|_2,S_2|_{-1}\]+\[M_3|_0,s_1|_{1}\]+\[m_3'|_2,S_1|_{-1}\],\nonumber\\
\lineup =\[Q,s_3|_{1}\]+\[M_2|_0,s_2^{(\mathrm{I})}|_1\]+\[M_3|_0,s_1|_{1}\]+\[M_2|_0,s_2^{(\mathrm{II})}|_1\]+\[m_2|_2,S_2|_{-1}\]+\[m_3'|_2,S_1|_{-1}\],\nonumber\\
\lineup = \[Q,s_3|_{1}\]+\[M_2|_0,s_2^{(\mathrm{I})}|_1\]+\frac{1}{2}\[\[Q,\mu_3|_0\],s_1|_{1}\]+\frac{1}{2}\[\[M_2|_0,\mu_2|_0\],s_1|_{1}\],\nonumber\\
\lineup\ \ \ +\[M_2|_0,s_2^{(\mathrm{II})}|_1\]+\[m_2|_2,\[S_1|_{-1},\mu_2|_0\]\]+\[\[m_2|_2,\mu_2|_0\],S_1|_{-1}\],\nonumber\\
\lineup = \[Q,s_3|_{1}\]+\[M_2|_0,s_2^{(\mathrm{I})}|_1\]+\frac{1}{2}\[Q,\[\mu_3|_0,s_1|_{1}\]\]-\frac{1}{2}\[M_2|_0,s_2^{(\mathrm{I})}|_{1}\]+\frac{1}{2}\[\[M_2|_0,s_1|_1\],\mu_2|_0\],\nonumber\\
\lineup\ \ \ +\[M_2|_0,s_2^{(\mathrm{II})}|_1\]+\[\[m_2|_2,S_1|_{-1}\],\mu_2|_0\],\nonumber\\
\lineup = \[Q,s_3|_{1}\]+\frac{1}{2}\[Q,\[\mu_3|_0,s_1|_{1}\]\]+\frac{1}{2}\[M_2|_0,s_2^{(\mathrm{I})}|_{1}\]-\frac{1}{2}\[\[Q,s_2^{(\mathrm{I})}|_1\],\mu_2|_0\],\nonumber\\
\lineup\ \ \ +\[M_2|_0,s_2^{(\mathrm{II})}|_1\]-\[\[Q,s_2^{(\mathrm{II})}|_1\],\mu_2|_0\],\nonumber\\
\lineup = \left[Q,s_3|_1 -\frac{1}{2}\Big(\[s_1|_{1},\mu_3|_0\]+\[s_2^{(\mathrm{I})}|_1,\mu_2|_0\]\Big)-\[s_2^{(\mathrm{II})}|_1,\mu_2|_0\]\right].
\end{eqnarray}
Therefore we can identify
\begin{equation}s_3|_1 = s_3^{(\mathrm{I})}|_1 + s_3^{(\mathrm{II})}|_1,\end{equation}
with
\begin{eqnarray}
s_3^{(\mathrm{I})}|_1 \lineup = \frac{1}{2}\Big(\[s_1|_{1},\mu_3|_0\]+\[s_2^{(\mathrm{I})}|_1,\mu_2|_0\]\Big),\\
s_3^{(\mathrm{II})}|_1\lineup = \[s_2^{(\mathrm{II})}|_1,\mu_2|_0\].
\end{eqnarray}
Thus we see a pattern where the supersymmetry product at Ramond number 1 breaks up into a product denoted with $(\mathrm{I})$ and a product denoted with $(\mathrm{II})$, each determined by independent recursions. Let us check that $s_3|_1$ is in the small Hilbert space:
\begin{eqnarray}
[\eta,s_3|_1]\lineup = \frac{1}{2}\Big(\[s_1|_{1},m_3|_0\]+\[s_2^{(\mathrm{I})}|_1,m_2|_0\]+\[\[s_1|_1,m_2|_0\],\mu_2|_0\]\Big)+\[s_2^{(\mathrm{II})}|_1,m_2|_0\]-\[\[m_2|_2,s_1|_{-1}\],\mu_2|_0\],\nonumber\\
\lineup = \[\[s_1|_1,m_2|_0\],\mu_2|_0\]+\[\[m_2|_2,\sigma_1|_{-1}\],m_2|_0\]-\[\[m_2|_2,s_1|_{-1}\],\mu_2|_0\],\nonumber\\
\lineup = \[[s_1,m_2]|_1,\mu_2|_0\]-\frac{1}{2}\[[m_2,m_2]|_2,\sigma_1|_{-1}\]+\[m_2|_2,[\sigma_1,m_2]|_{-1}\],\nonumber\\
\lineup = 0.
\end{eqnarray}
This completes the definition of the 3-product $\tilde{S}_3$ in the supersymmetry transformation.

Now we can guess the form of the supersymmetry transformation at higher orders. The $(n+1)$st product can be written
\begin{equation}\tilde{S}_{n+1} = S_{n+1}|_{-1} + s_{n+1}|_1.\end{equation}
Components with higher Ramond number can be set to zero. In addition, $s_{n+2}|_1$ can be written as a sum
\begin{equation}s_{n+2}|_1 = s_{n+2}^{(\mathrm{I})}|_1 + s_{n+2}^{(\mathrm{II})}|_1.\end{equation}
The products are determined recursively by the equations,
\begin{eqnarray}
S_{n+2}|_{-1} \lineup = \frac{1}{n+1}\sum_{k=0}^n\[S_{k+1}|_{-1},\mu_{n-k+2}|_0\],\label{eq:recS}\\
s_{n+2}^{(\mathrm{I})}|_{1} \lineup = \frac{1}{n+1}\sum_{k=0}^n\[s_{k+1}^{(\mathrm{I})}|_{1},\mu_{n-k+2}|_0\],\\
s_{n+3}^{(\mathrm{II})}|_{1} \lineup = \frac{1}{n+1}\sum_{k=0}^n\[s_{k+2}^{(\mathrm{II})}|_{1},\mu_{n-k+2}|_0\],\label{eq:recspp}
\end{eqnarray}
starting from $S_1|_{-1}$ in \eq{S1}, $s_1^{(\mathrm{I})}|_1 = s_1|_1$ in \eq{s1}, and $s_2^{(\mathrm{II})}|_1 = \[m_2|_2,\sigma_1|_{-1}\]$. Now let's prove that these products have the required properties. We promote the products to coderivations and define generating functions
\begin{eqnarray}
\S(t)\lineup = \sum_{n=0}^{\infty} t^n\S_{n+1}|_{-1},\nonumber\\
\s^{(\mathrm{I})}(t)\lineup = \sum_{n=0}^{\infty} t^n \s_{n+1}^{(\mathrm{I})}|_1,\nonumber\\
\s^{(\mathrm{II})}(t)\lineup = \sum_{n=0}^{\infty} t^n \s_{n+2}^{(\mathrm{II})}|_1.
\end{eqnarray}
Using the generating function of the gauge products \eq{mugen}, the recursive equations \eq{recS}-\eq{recspp} can be reexpressed
\begin{eqnarray}
\frac{d}{dt} \S(t)\lineup = [\S(t),\mmu(t)],\label{eq:Sdiff}\\
\frac{d}{dt} \s^{(\mathrm{I})}(t)\lineup = [\s^{(\mathrm{I})}(t),\mmu(t)],\\
\frac{d}{dt} \s^{(\mathrm{II})}(t)\lineup = [\s^{(\mathrm{II})}(t),\mmu(t)],\label{eq:S2diff}
\end{eqnarray}
The generating functions for the products in the equations of motion and supersymmetry transformation take the form
\begin{equation}\tilde{\M}(t) = \M(t) + t\m'(t),\ \ \ \ \tS(t) = \S(t) + \s^{(\mathrm{I})}(t) + t\s^{(\mathrm{II})}(t).\end{equation}
The differential equations for the generating functions imply a set of equations:
\begin{eqnarray}
\frac{d}{dt} [\tilde{\M}(t),\tilde{\S}(t)] \lineup = [[\tilde{\M}(t),\tilde{\S}(t)] ,\mmu(t)]+ [\m'(t),\tS(t)] + [\tilde{\M}(t),\s^{(\mathrm{II})}(t)],\label{eq:Sfirst}\\
\frac{d}{dt}\Big([\m'(t),\tS(t)] + [\tilde{\M}(t),\s^{(\mathrm{II})}(t)]\Big) \lineup = [[\m'(t),\tS(t)] + [\tilde{\M}(t),\s^{(\mathrm{II})}(t)],\mmu(t)] + 2[\m'(t),\s^{(\mathrm{II})}(t)],\label{eq:Sntl}\\
\frac{d}{dt}[\m'(t),\s^{(\mathrm{II})}(t)] \lineup = [[\m'(t),\s^{(\mathrm{II})}(t)],\mmu(t)].
\end{eqnarray}
Start with the last equation. Note that $[\m'(0),\s^{(\mathrm{II})}(0)]$ vanishes because 
\begin{equation}[m_2|_2,s_2^{(\mathrm{II})}|_1] =[m_2|_2, [m_2|_2,\sigma_1|_{-1}]]=0.\end{equation}
The last equation then implies 
\begin{equation}[\m'(t),\s^{(\mathrm{II})}(t)] = 0.\end{equation}
The next to last equation \eq{Sntl} is now homogeneous in $[\m'(t),\tS(t)] + [\tilde{\M}(t),\s^{(\mathrm{II})}(t)]$. We know that $[\m'(0),\tS(0)] + [\tilde{\M}(0),\s^{(\mathrm{II})}(0)]=0$ because 
\begin{equation}[m_2|_2,S_1|_{-1}] + [Q,[m_2|_2,\sigma_1|_{-1}]] = 0.\end{equation}
Therefore \eq{Sntl} implies
\begin{equation}[\m'(t),\tS(t)] + [\tilde{\M}(t),\s^{(\mathrm{II})}(t)]=0.\end{equation}
Finally, consider the first equation \eq{Sfirst}, which is now homogeneous in $[\tilde{\M}(t),\tS(t)]$. The commutator $[\tilde{\M}(0),\tilde{\S}(0)]$ vanishes  since the supersymmetry operators are BRST invariant. Therefore
\begin{equation}[\tilde{\M}(t),\tS(t)]=0.\end{equation}
Setting $t=1$, we have in particular 
\begin{equation}[\tilde{\M},\tS]=0,\end{equation}
which proves that the products $\tilde{S}_{n+1}$ generate a symmetry. To prove that the transformation acts in the small Hilbert space, consider the following set of equations:
\begin{eqnarray}
\frac{d}{dt}[\n,\tS(t)]\lineup = [[\n,\tS(t)],\mmu(t)] + [\tS(t),\m(t)] + [\n,\s^{(\mathrm{II})}(t)],\label{eq:nSfirst}\\
\frac{d}{dt}\Big([\tS(t),\m(t)] + [\n,\s^{(\mathrm{II})}(t)]\Big)\lineup =[[\tS(t),\m(t)] + [\n,\s^{(\mathrm{II})}(t)],\mmu(t)] + 2[\s^{(\mathrm{II})}(t),\m(t)],\label{eq:nSntl}\\
\frac{d}{dt}[\s^{(\mathrm{II})}(t),\m(t)]\lineup = [[\s^{(\mathrm{II})}(t),\m(t)],\mmu(t)].
\end{eqnarray}
Start with the last equation. Note that $[\m(0),\s^{(\mathrm{II})}(0)]$ vanishes because 
\begin{equation}[m_2|_0,s_2^{(\mathrm{II})}|_1] =[m_2|_0, [m_2|_2,\sigma_1|_{-1}]]=\frac{1}{2}[[m_2, m_2]|_2,\sigma_1|_{-1}]-[m_2|_2,[m_2,\sigma_1]|_{-1}]=0.\end{equation} 
The last equation then implies 
\begin{equation}[\m(t),\s^{(\mathrm{II})}(t)] = 0.\end{equation}
The next to last equation \eq{nSntl} is now homogeneous in $[\tS(t),\m(t)] + [\n,\s^{(\mathrm{II})}(t)]$. We know that $[\tS(0),\m(0)] + [\n,\s^{(\mathrm{II})}(0)]=0$ because 
\begin{equation}[S_1|_{-1}+s_1|_1,m_2|_0]+[\eta,[m_2|_2,\sigma_1|_{-1}] = [S_1,m_2]|_0+[s_1,m_2]|_1 = 0.\end{equation}
Therefore \eq{nSntl} implies
\begin{equation}
[\tS(t),\m(t)] + [\n,\s^{(\mathrm{II})}(t)]=0.
\end{equation}
Finally, consider the first equation \eq{nSfirst}, which is now homogeneous in $[\n,\tS(t)]$. Since $[\n,\tilde{\S}(0)] =0$ we conclude
\begin{equation}[\n,\tS(t)]=0,\end{equation}
so the products $\tilde{S}_{n+1}$ are in the small Hilbert space. This completes the construction of the supersymmetry transformation.

\subsection{Supersymmetry in Berkovits' Superstring Field Theory}

Following \cite{OkWB,WB}, our analysis should imply a natural form for the supersymmetry transformation of the NS+R equations of motion in Berkovits' open superstring field theory \cite{BerkRamond,Berkovits1,Berkovits2}. To derive this supersymmetry transformation, as an intermediate step we must perform a field redefinition from our string field $\tilde{\Phi}$ to a new string field $\tilde{\varphi}$ whose equations of motion and constraint are only polynomial \cite{WB}. The field redefinition is defined by the cohomomorphism
\begin{equation}\G(t) = \mathcal{P}\left[\exp\left(\int_0^t dt'\, \mmu(t')\right)\right],\end{equation}
where the path ordered exponential is defined in sequence of increasing $t'$. In particular, we can express the relation between $\tilde{\Phi}$ and $\tilde{\varphi}$ as 
\begin{equation}\frac{1}{1-\tilde{\varphi}} = \G \frac{1}{1-\tilde{\Phi}},\end{equation}
where $\G\equiv\G(1)$. Equivalently, we can write
\begin{equation}\tilde{\varphi} = G[\tilde{\Phi}],\label{eq:vartphi}\end{equation}
where $G[\tilde{\Phi}]$ is given by
\begin{eqnarray}G[\tilde{\Phi}]\lineup  \equiv \pi_1\G\frac{1}{1-\tilde{\Phi}},\nonumber\\
\lineup = \tilde{\Phi} + \mu_2|_0(\tilde{\Phi},\tilde{\Phi})+ \frac{1}{2}\Big(\mu_3|_0(\tilde{\Phi},\tilde{\Phi},\tilde{\Phi})+\mu_2|_0(\mu_2|_0(\tilde{\Phi},\tilde{\Phi}),\tilde{\Phi}) + \mu_2|_0(\tilde{\Phi},\mu_2|_0(\tilde{\Phi},\tilde{\Phi}))\Big)+\mathrm{higher\ orders},\ \ \ \ \ \label{eq:G}\end{eqnarray}
and $\pi_1$ denotes the projection onto the 1-string component of the tensor algebra. Note that this field redefinition is only defined in the large Hilbert space---it is not compatible with the small Hilbert space constraint on the string field. In \cite{WB}, such a field redefinition was called an 
{\it improper} field redefinition. Once we have transformed from $\tilde{\Phi}$ to $\tilde{\varphi}$, we must perform a second transformation to map from $\tilde{\varphi}$ into Berkovits' superstring field theory. However, it will be useful to work out the supersymmetry transformation for $\tilde{\varphi}$ first. 

Let us find the equations of motion and constraint for $\tilde{\varphi}$. The generating functions given in \eq{Mgen}-\eq{mgen} can be written
\begin{eqnarray}
\M(t)\lineup = \G(t)^{-1}\Q\G(t),\\
\m'(t)\lineup = \G(t)^{-1}\m_2|_2\G(t),\\
\m(t)\lineup = \G(t)^{-1}\m_2|_0\G(t) ,
\end{eqnarray}
since these expressions give a solution to the differential equations \eq{Mdiff}-\eq{mdiff} with the correct initial conditions. This implies that the composite products of the equations of motion can be expressed:
\begin{equation}\tilde{\M} = \M(1) + \m'(1) =  \G^{-1}\Big(\Q + \m_2|_2\Big)\G.\label{eq:tMtrans}\end{equation}
Moreover, it follows from the computation in \cite{OkWB} that 
\begin{equation}\n = \G^{-1}\Big(\n - \m_2|_0\Big)\G.\label{eq:ntrans}\end{equation}
The equations of motion and constraint for $\tilde{\Phi}$ can be expressed together in the form
\begin{equation}\Big(\tilde{\M}-\n\Big)\frac{1}{1-\tilde{\Phi}} = 0.\end{equation}
Multiplying by $\G$, using \eq{tMtrans} and \eq{ntrans}, and replacing $\tilde{\Phi}$ with the new string field $\tilde{\varphi}$, we find:
\begin{equation}\Big(\Q-\n +\m_2\Big)\frac{1}{1-\tilde{\varphi}} = 0.\label{eq:polytensorEOM}\end{equation}
After projecting onto the 1-string component of the tensor algebra, we find familiar Chern-Simons-like equations
\begin{equation}(Q-\eta)\tilde{\varphi}+ \tilde{\varphi}*\tilde{\varphi} = 0.\label{eq:poly}\end{equation}
The field redefinition implies that we can break $\tilde{\varphi}$ into NS and Ramond components  
\begin{equation}\tilde{\varphi} =\varphi_\mathrm{N} + \psi_\mathrm{R},\end{equation}
where the NS field $\varphi_\mathrm{N}$ is Grassmann odd and has picture $-1$ and the R field $\psi_\mathrm{R}$ is Grassmann odd and has picture $-1/2$. Then \eq{poly} is equivalent to four equations:
\begin{eqnarray}
0\lineup = Q\psi_\mathrm{R},\label{eq:polyEOM1}\\
0\lineup = Q\varphi_\mathrm{N} +\psi_\mathrm{R}*\psi_\mathrm{R},\label{eq:polyEOM2}\\
0\lineup = \eta \psi_\mathrm{R} - [\psi_\mathrm{R},\varphi_\mathrm{N}],\label{eq:polyconst1}\\
0\lineup = \eta \varphi_\mathrm{N} - \varphi_\mathrm{N}*\varphi_\mathrm{N},\label{eq:polyconst2}
\end{eqnarray}
where the commutator of string fields is computed with the star product and is graded with respect to Grassmann parity. The first two equations should be interpreted as equations of motion, since they are the image of the equations of motion for $\tilde{\Phi}$ after the field redefinition. The second two equations should be interpreted as equations of constraint, since they are the image of the small Hilbert space constraint on $\tilde{\Phi}$ after the field redefinition. Note that the equations of motion for $\tilde{\varphi}$ are quadratic the the Ramond string field, while the equations of motion for $\tilde{\Phi}$ are cubic in the Ramond field. This happens because the field redefinition \eq{G} is at most linear in the Ramond field, and, in conjunction with the quadratic term in the Ramond field in the equations of motion for $\tilde{\varphi}$, this produces cubic terms in the Ramond field in the equations of motion for $\tilde{\Phi}$.

Now let's derive the new form of the supersymmetry transformation. The differential equations \eq{Sdiff}-\eq{S2diff} imply that the supersymmetry transformation can be written in the form
\begin{equation}
\tilde{\S} = \G^{-1}\q\G.
\end{equation}
where
\begin{equation}\q\equiv \S_1|_{-1}+\s_1|_1 + [\ssigma_1|_{-1},\m_2|_2].\label{eq:q}\end{equation}
This implies that the supersymmetry transformation of $\tilde{\varphi}$ can be expressed 
\begin{equation}\delta \frac{1}{1-\tilde{\varphi}} = \q\frac{1}{1-\tilde{\varphi}}.\end{equation}
Explicitly in terms of the NS and R components, 
\begin{eqnarray}
\delta\varphi_\mathrm{N} \lineup =s_1\psi_\mathrm{R}+[\psi_\mathrm{R},\sigma_1\phi_\mathrm{N}],\label{eq:polyNSSUSY}\\
\delta\psi_\mathrm{R} \lineup = S_1\varphi_\mathrm{N} + \sigma_1(\psi_\mathrm{R}*\psi_\mathrm{R}).\label{eq:polyRSUSY}
\end{eqnarray}
We can check that $\q$ is a symmetry of the equations of motion:  
\begin{eqnarray}
\[\q,\Q+\m_2|_2\] \lineup = \[\S_1|_{-1}+\s_1|_1 + [\ssigma_1|_{-1},\m_2|_2],\Q+\m_2|_2\],\nonumber\\
\lineup = -\[\S_1|_{-1},\m_2|_2\] + \[\S_1|_{-1},\m_2|_2\]+\[\s_1|_{1},\m_2|_2\]+\[\[\ssigma_1|_{-1},\m_2|_2\],\m_2|_2\],\nonumber\\
\lineup = 0.
\end{eqnarray}
The terms either cancel or vanish because the Ramond number exceeds the number of inputs. We can also check that $\q$ preserves the constraints:
\begin{eqnarray}
\[\q,\n-\m_2|_0\] \lineup = \[\S_1|_{-1}+\s_1|_1 + \[\ssigma_1|_{-1},\m_2|_2\],\n-\m_2|_0\],\nonumber\\
\lineup = -\[\s_1|_{-1},\m_2|_2\] - \[\S_1|_{-1},\m_2|_0\]-\[\s_1|_{1},\m_2|_0\]-\[\[\ssigma_1|_{-1},\m_2|_2\],\m_2|_0\],\nonumber\\
\lineup = - \[\s_1,\m_2\]|_1-\[\S_1,\m_2\]|_{-1}-\frac{1}{2}\[\ssigma_1|_{-1},\[\m_2,\m_2]|_2\]+\[\[\ssigma_1,\m_2\]|_{-1},\m_2|_2\],\nonumber\\
\lineup = 0.
\end{eqnarray}
This vanishes since the $s_1,S_1$ and $\sigma_1$ are derivations of the star product and because the star product is associative. Conjugating by $\G$, this provides an alternative proof that the supersymmetry transformation $\tilde{\S}$ preserves the equations of motion for $\tilde{\Phi}$ and is consistent with the small Hilbert space constraint. 

Now that we have the supersymmetry transformation for $\tilde{\varphi}$, it is straightforward to translate into Berkovits' superstring field theory. Berkovits' superstring field theory uses a Grassmann even NS field $\Phi$ in the large Hilbert space and at picture and ghost number $0$, and a Grassmann odd R string field $\Omega$ in the small Hilbert space at picture $-1/2$ and at ghost number $1$ \cite{BerkRamond}. These are related to $\varphi_{\mathrm{N}}$ and $\psi_{\mathrm{R}}$ through\footnote{The string field we call $\Omega$ here is called $i\Omega$ in \cite{BerkRamond}.}
\begin{equation}\varphi_{\mathrm{N}} = (\eta e^\Phi)e^{-\Phi},\ \ \ \ \psi_\mathrm{R} = e^{\Phi}\Omega e^{-\Phi} .\label{eq:polyBerktrans}\end{equation}
With this identification, the constraints \eq{polyconst1} and \eq{polyconst2} reduce to identities, and the equations of motion \eq{polyEOM1} and \eq{polyEOM2} translate to
\begin{equation}
\eta\Big(e^{-\Phi} Qe^{\Phi}\Big) = \Omega^2,\ \ \ \ \ Q\Omega +\[e^{-\Phi}Qe^{\Phi},\Omega\]= 0.
\end{equation}
The Berkovits equations of motion are invariant under infinitesimal gauge invariances
\begin{eqnarray}\delta e^{\Phi} \lineup = e^{\Phi}(v + [\Omega,\Lambda]) + ue^{\Phi},\\
\delta\Omega\lineup = [\Omega,v] +\eta\Big(Q\Lambda + \[e^{-\Phi}Qe^{\Phi},\Lambda\]\Big).
\end{eqnarray}
The NS gauge parameters $u,v$ are Grassmann even and carry ghost and picture number zero; $u$ is BRST closed while $v$ is in the small Hilbert space. The Ramond gauge parameter $\Lambda$ is Grassmann odd and in the large Hilbert space, and carries ghost number $-1$ and picture number $+1/2$. The supersymmetry transformation is
\begin{equation}\delta\Omega = \eta\Big(e^{-\Phi}S_1e^{\Phi}\Big),\ \ \ \ (\delta e^{\Phi})e^{-\Phi}= \sigma_1\Big(e^{\Phi}\Omega e^{-\Phi}\Big).\label{eq:BerkSUSY}\end{equation}
Plugging into \eq{polyBerktrans} gives the supersymmetry transformation $\q$ for the string field $\tilde{\varphi}$, and further mapping gives the original supersymmetry transformation $\tilde{\S}$ for the string field $\tilde{\Phi}$.

One interesting potential application of these results is to check whether or not classical solutions in open superstring field theory are supersymmetric. However, the supersymmetry transformation we have derived is not necessarily the most convenient for this purpose. Specifically, our supersymmetry transformation is natural for an NS string field at picture $-1$, but usually analytic solutions in Berkovits superstring field theory are closely related to an NS string field at picture $0$.\footnote{Often it is possible to ``dualize" analytic solutions into a form which is natural at picture $-1$ \cite{democraticgf}.}  For this reason, we will find it convenient to consider a ``dual" formulation of open superstring field theory using the string field 
\begin{equation}\tilde{\varphi}^* = \varphi^*_\mathrm{N} + \psi_\mathrm{R}^*,\end{equation}
where $\varphi^*_\mathrm{N}$ is a Grassmann odd NS field at picture 0 and ghost number 1 and $\psi^*_\mathrm{R}$ is a Grassmann odd Ramond field at picture $-1/2$ and ghost number 1. The ``dual" fields satisfy the equations
\begin{eqnarray}
0\lineup =Q\varphi^*_\mathrm{N}+\varphi^*_\mathrm{N}*\varphi^*_\mathrm{N}, \label{eq:modconst1} \\
0\lineup = Q\psi^*_\mathrm{R} + [\varphi^*_\mathrm{N},\psi^*_\mathrm{R}],\label{eq:modconst2}\\
0\lineup = \eta\varphi^*_{\mathrm{N}}-\psi^*_\mathrm{R}*\psi^*_\mathrm{R},\label{eq:modEOM1}\\
0\lineup = \eta\psi^*_\mathrm{R},\label{eq:modEOM2}
\end{eqnarray}
When $\psi_\mathrm{R}^*=0$, these are the equations of motion of the NS sector of the modified cubic superstring field theory \cite{PTY,Russians}. Note that the new field equations for $\tilde{\varphi}^*$ are identical to the previous field equations for $\tilde{\varphi}$ after the ``duality" map: 
\begin{equation}Q\leftrightarrow -\eta,\ \ \ \ \mathrm{picture}\leftrightarrow-(1+\mathrm{picture}).\end{equation}
We indicate the ``dual" fields with a star $*$. We have
\begin{equation}
0=\Big(\Q-\n+\m_2\Big)\frac{1}{1-\tilde{\varphi}^*}.
\end{equation}
Taking the ``dual" of the supersymmetry transformation for $\tilde{\varphi}$ we arrive at a new supersymmetry transformation for $\tilde{\varphi}^*$
\begin{equation}
\delta^*\frac{1}{1-\tilde{\varphi}^*} = \q^*\frac{1}{1-\tilde{\varphi}^*},
\end{equation}
where
\begin{equation}\q^*\equiv \s_1|_{-1} + \S_1|_1 -\[\ssigma_1|_{-1},\m_2|_2\] .\label{eq:Stpi}\end{equation}
Relating this ``dual" formulation to Berkovits' superstring field theory requires a slight change of perspective on the Ramond sector. In the previous paragraph we described the Ramond sector of the Berkovits theory using a string field $\Omega$ in the small Hilbert space. Now it will be more natural to use a string field $\Omega^*$ which is not in the small Hilbert space but is BRST closed. The Ramond field $\Omega^*$ is Grassmann odd and carries ghost number 1 and picture $-1/2$.  On shell, the Ramond fields $\Omega$ and $\Omega^*$ are related by
\begin{equation}\Omega = e^{-\Phi}\Omega^* e^{\Phi},\ \ \ (\mathrm{on\ shell}).\end{equation}
This relation is not meaningful off-shell since it is not consistent with the assumption that $\Omega^*$ is BRST closed while $\Omega$ is in the small Hilbert space. The ``dual" fields $\varphi^*_\mathrm{N}$ and $\psi^*_\mathrm{R}$ can be then be related to Berkovits' superstring field theory through
\begin{equation}\varphi^*_\mathrm{N} = e^{-\Phi}Qe^{\Phi},\ \ \ \psi^*_\mathrm{R} = e^{-\Phi}\Omega^* e^{\Phi}\label{eq:modBerk}\end{equation}
In this way, \eq{modconst1} and \eq{modconst2} reduce to identities while \eq{modEOM1} and \eq{modEOM2} are equivalent to the equations of motion of Berkovits' superstring field theory expressed in the form
\begin{equation}Q\Big((\eta e^{\Phi})e^{-\Phi}\Big) +(\Omega^*)^2 = 0,\ \ \ \ \eta\Omega^* - \[(\eta e^{\Phi})e^{-\Phi},\Omega^*\] = 0.\end{equation}
In these variables, the ``dual" supersymmetry transformation of the Berkovits theory takes the form
\begin{equation}\delta^*\Omega^* = Q\Big((s_1 e^{\Phi})e^{-\Phi}\Big),\ \ \ \ e^{-\Phi}\delta^* e^\Phi=\sigma_1\Big(e^{-\Phi}\Omega^* e^{\Phi}\Big).\end{equation}
Plugging into \eq{modBerk} produces to the supersymmetry transformation of $\varphi^*_\mathrm{N}$ and $\psi^*_\mathrm{R}$ given in \eq{Stpi}. 

Note that the supersymmetry transformation $\delta$ and the ``dual" supersymmetry transformation $\delta^*$ assume different off-shell degrees of freedom, and therefore really apply to different string field theories. On shell, however, they should be equivalent. In fact, one can show that they are equal up to an infinitesimal gauge transformation:
\begin{eqnarray}
\delta^*e^{\Phi} \lineup= \delta e^{\Phi} + e^{\Phi}[\Omega,\Lambda], \ \ \ \ \ \ \ \ \ \ \ \ \ \ \ \ \ \ \ \ \ \ (\mathrm{on\ shell}),\\  
\delta^*\Omega \lineup = \delta\Omega +\eta\Big(Q\Lambda + \[e^{-\Phi}Qe^{\Phi},\Lambda\]\Big),\ \ \ \ (\mathrm{on\ shell}),\label{eq:fermsusy}
\end{eqnarray}
where the Ramond gauge parameter is given by
\begin{equation}\Lambda = - e^{-\Phi}\sigma_1e^{\Phi}.\end{equation}
This shows that the two supersymmetry transformations are physically equivalent.

We are now ready to discuss supersymmetry of classical solutions. On a BPS D-brane there are not many classical solutions whose existence is well-established. The only known solutions represent marginal deformations of the reference boundary superconformal field theory. We will consider specifically a transverse displacement of the reference D-brane, for which it is sufficient to consider the solution \cite{supermarg,Oksupermarg}
\begin{equation}e^\Phi = 1+ F X\frac{1}{1+B\frac{1-F^2}{K}J} F,\end{equation}
in the context of Berkovits' superstring field theory. As is appropriate for a classical background, the Ramond string field is zero. We use the algebraic notation for analytic solutions introduced in \cite{Okawa}, following the conventions of \cite{simple}. The string field $F$ is a function of $K$ which we can take to be the square root of the $SL(2,\mathbb{R})$ vacuum.\footnote{We use the algebraic notation introduced in \cite{Okawa}. See also \cite{simple} for the specific conventions that we follow.} The fields $X$ and $J$ are defined
\begin{equation}X \equiv \lambda\, \xi e^{-\phi}c\psi^+,\ \ \ \ J \equiv QX = \lambda\, \big(i\sqrt{2}c\d X^+ +\gamma\psi^+\big),\end{equation}
where $\lambda$ is the marginal parameter describing the displacement of the D-brane, and the index $+$ indicates a lightcone direction whose spatial component is transverse to the D-brane.\footnote{Technically, this solution translates the D-brane and switches on a timelike Wilson line of corresponding magnitude. However, as described in \cite{KOSing}, the timelike Wilson line is physically trivial. An alternative approach would use the solution of \cite{KOsuper}, which does not require excitation of spurious primaries in the timelike factor of the $X^{\mu}$ and $\psi^\mu$ BCFTs.} The Berkovits solution implies an expression for the NS field $\varphi_\mathrm{N}$ at picture $-1$ and the ``dual" NS field $\varphi^*_\mathrm{N}$ at picture zero:
\begin{eqnarray}
\varphi_\mathrm{N} \lineup = (\eta e^{\Phi})e^{-\Phi} = F(\eta X)\frac{1}{1+B\frac{1-F^2}{K}J+F^2 X}F,\label{eq:varphiNsol}\\
\varphi^*_\mathrm{N} \lineup = e^{-\Phi}Qe^{\Phi} = F J\frac{1}{1+B\frac{1-F^2}{K}J} F.\label{eq:varphistNsol}
\end{eqnarray}
Note that the picture zero field is simpler, which is why the ``dual" supersymmetry transformation $\delta^*$ is more convenient. The Berkovits solution also implies a solution $\tilde{\Phi}$ of the original equations of motion in the small Hilbert space, which up to second order in the marginal operator takes the form
\begin{equation}
\Phi_\mathrm{N} = F (\eta X) F - F(\eta X) B\frac{1-F^2}{K} JF - F (\eta X) F^2 X F -\mu_2\Big(F(\eta X)F,F(\eta X)F\Big) +\mathrm{higher\ orders}.
\label{eq:smalltrans}\end{equation}
Translation of a D-brane does not break any supersymmetries. Therefore we should be able to show that this solution is supersymmetric. The only nontrivial supersymmetry transformation to compute is for the fermion.  Using the ``dual" supersymmetry transformation we find
\begin{eqnarray}\delta^*\Omega^* \lineup = Q\Big((s_1e^{\Phi})e^{-\Phi}\Big),\nonumber\\
\lineup = e^{\Phi}s_1(e^{-\Phi}Qe^\Phi)e^{-\Phi},\nonumber\\
\lineup = e^{\Phi} (s_1 \varphi^*_\mathrm{N})e^{-\Phi}.
\end{eqnarray} 
Thus we need to compute $s_1\tilde{\varphi}^*_\mathrm{N}$. This is zero because $s_1$ is a derivation of the star product and because
\begin{equation}s_1 K = 0,\ \ \ s_1 B=0,\ \ \ s_1 J = 0.\end{equation}
The first equation follows because $s_1$ is the zero mode of a weight 1 primary, and the last two equations follow because there are no poles in the OPE between the $-1/2$ picture fermion vertex and either $b$ or $J$. Therefore the solution is supersymmetric. However, the solution is not identically invariant under the supersymmetry transformation $\delta$ which is natural at picture $-1$. From \eq{fermsusy} one finds that the supersymmetry transformation of the fermion is
\begin{equation}\delta\Omega = \eta\Big(Q\Lambda + \[e^{-\Phi}Qe^{\Phi},\Lambda\]\Big).\end{equation}
with $\Lambda = -e^{-\Phi}\sigma_1 e^{\Phi}$. While $\sigma_1$ annihilates $K$ and $B$, it does not annihilate $J$, and the Ramond gauge parameter $\Lambda$ is not zero. Translating back into our original equations of motion, this means that the supersymmetry transformation $\tilde{\S}$ constructed in the previous section leaves the solution $\tilde{\Phi}$ in \eq{smalltrans} invariant up to a gauge transformation. 

While it is interesting to know the supersymmetry transformation on a BPS D-brane, it is not really enough to give a full account of the role of supersymmetry in open superstring field theory. This is because at least half of the supersymmetries are spontaneously broken by the reference boundary superconformal field theory, and broken supersymmetries can in principle be restored upon expanding around a nontrivial classical solution. The most dramatic example of this is tachyon condensation on a non-BPS D-brane \cite{supervac}, where all 32 supersymmetries are restored at the tachyon vacuum. This can be seen as follows. While we do not presently know the explicit form of broken supersymmetry transformations in superstring field theory, in any case they will generate at most an infinitesimal deformation of the tachyon vacuum.  But since the kinetic operator around the tachyon vacuum has no cohomology, any infinitesimal deformation can be removed by gauge transformation, which shows that the tachyon vacuum is invariant under all supersymmetries. (In fact, this argument shows that the tachyon vacuum is invariant under all symmetries of the closed string background). For more general BPS solutions the explicit realization of this story will be more nontrivial, but progress along these lines may be possible following \cite{KOSing}.

\subsection{Supersymmetry Algebra}

Now that we have a supersymmetry transformation, it is interesting to compute the supersymmetry algebra. For this purpose it is easier to work with the polynomial string field $\tilde{\varphi}$ in \eq{vartphi} rather than the original string field $\tilde{\Phi}$ in the small Hilbert space. Since the supersymmetry transformation of $\tilde{\varphi}$ is given by the coderivation $\q$, we compute
\begin{equation}[\q,\q'],\end{equation}
where the prime indicates that the operator is defined with a second independent supersymmetry parameter $\eps_{\vec{s}}'$. Once we find $\[\q,\q']$, we can easily recover $[\tilde{\S},\tilde{\S}']$ for the original string field $\tilde{\Phi}$ by making a similarity transformation with $\G$. 

The commutator of supersymmetry transformations should produce the momentum operator: 
\begin{equation}P_\mu\equiv \oint_{|z|=1}\frac{dz}{2\pi i} i\d X_\mu(z).\end{equation} 
It will also be useful to consider a ``momentum operator" at picture $-1$
\begin{equation}
p_\mu\equiv \frac{1}{\sqrt{2}}\oint_{|z|=1}\frac{dz}{2\pi i} \psi_\mu e^{-\phi}(z),
\end{equation}
and
\begin{equation}
\pi_\mu \equiv \frac{1}{\sqrt{2}}\oint_{|z|=1}\frac{dz}{2\pi i} \xi\psi_\mu e^{-\phi}(z).
\end{equation}
The operators $P_\mu,\pi_\mu$ and $p_\mu$ are all derivations of the star product. Moreover, they satisfy
\begin{eqnarray}
P_\mu \lineup = [Q,\pi_\mu],\\
p_\mu \lineup = [\eta,\pi_\mu],
\end{eqnarray}
analogous to the relation between $M_2,\mu_2$ and $m_2$. To compute the supersymmetry algebra we need the following commutators between supersymmetry operators:
\begin{eqnarray}[s_1,s_1'] \lineup = -2 \, p(\eps,\eps'),\label{eq:ssp}\\
\ [S_1,s_1'] = [s_1,S_1'] \lineup = -2\, P(\eps,\eps'),
\end{eqnarray}
where for short we denote
\begin{equation}
P(\eps,\eps')\equiv \big(\eps_{\vec{s}} \ C_{\vec{s},\vec{s}'}\Gamma^\mu_{\vec{s}',\vec{s}''}\, \eps'_{\vec{s}''}\big) P_\mu,
\end{equation}
and similarly for $p(\eps,\eps')$ and $\pi(\eps,\eps')$, where $C$ is the charge conjugation matrix.

Now we are ready to compute the supersymmetry algebra. Plugging in \eq{q} and expanding cross-terms gives
\begin{eqnarray}[\q,\q'] \lineup = \[\S_1|_{-1},\s'_1|_1\] + \[\s_1|_1,\S'_1|_{-1}\]+\[\S_1|_{-1},\[\ssigma'_1|_{-1},\m_2|_2\]\]+\[\[\ssigma_1|_{-1},\m_2|_2\],\S'_1|_{-1}\]\nonumber\\
\lineup \ \ \ +\[\s_1|_{1},\[\ssigma'_1|_{-1},\m_2|_2\]\]+\[\[\ssigma_1|_{-1},\m_2|_2\],\s'_1|_{1}\] +\[\[\ssigma_1|_{-1},\m_2|_2\],\[\ssigma'_1|_{-1},\m_2|_2\]\].\end{eqnarray}
To extract the momentum operator, rewrite the first two terms
\begin{eqnarray}
 \[\S_1|_{-1},\s'_1|_1\] + \[\s_1|_1,\S'_1|_{-1}\]\lineup =  [\S_1,\s'_1] + [\s_1,\S'_1]-\[\S_1|_{1},\s'_1|_{-1}\] - \[\s_1|_{-1},\S'_1|_{1}\],\nonumber\\
 \lineup = -4\,\P(\eps,\eps')-\[\S_1|_{1},\s'_1|_{-1}\] - \[\s_1|_{-1},\S'_1|_{1}\],\nonumber\\
  \lineup = -2\,\P(\eps,\eps')-2[\Q,\ppi(\eps,\eps')]-\[\S_1|_{1},\s'_1|_{-1}\] - \[\s_1|_{-1},\S'_1|_{1}\],
\end{eqnarray}
where $\P(\eps,\eps')$ is the coderivation corresponding to $P(\eps,\eps')$ and $\ppi(\eps,\eps')$ is the coderivation corresponding to $\pi(\eps,\eps')$. In the third step we chose to express part of the translation operator in the form $[Q,\pi_\mu]$, for reasons that will be clear shortly. Substituting we find 
\begin{eqnarray}[\q,\q'] \lineup = -2\,\P(\eps,\eps')-2[\Q,\ppi(\eps,\eps')]-\[\S_1|_{1},\s'_1|_{-1}\] - \[\s_1|_{-1},\S'_1|_{1}\]+\[\S_1|_{-1},\[\ssigma'_1|_{-1},\m_2|_2\]\]\nonumber\\ \lineup\ \ \ +\[\[\ssigma_1|_{-1},\m_2|_2\],\S'_1|_{-1}\]
+\[\s_1|_{1},\[\ssigma'_1|_{-1},\m_2|_2\]\]+\[\[\ssigma_1|_{-1},\m_2|_2\],\s'_1|_{1}\] +\[\[\ssigma_1|_{-1},\m_2|_2\],\[\ssigma'_1|_{-1},\m_2|_2\]\].\nonumber\\
\label{eq:step1}\end{eqnarray}
In the simplest supersymmetry algebra, the terms after $-2\,\P(\eps,\eps')$ would cancel. Unfortunately they do not cancel, and we have to make sense of them. The reason why the extra terms are present is that we are dealing with an on-shell supersymmetry algebra. This is not surprising, since the off-shell fermionic and bosonic degrees of freedom in the string field do not match. For example, at mass level $0$ we have 16 fermion fields but only 11 boson fields, including the gauge field, transverse scalars, and an auxiliary field. In the current context, on-shell supersymmetry implies that that the supersymmetry algebra should be expressible as
\begin{equation}\[\q,\q'\] = -2\,\P(\eps,\eps') + \[\Q+\m_2|_2,\Ll\],\label{eq:susyalg}\end{equation}
where $\Ll$ is a coderivation which is consistent with the constraint on the field $\tilde{\varphi}$:
\begin{equation}\[\n-\m_2|_0,\Ll\] = 0.\label{eq:Lconst}\end{equation}
Now we show that the supersymmetry algebra can be expressed in this form. Continuing from \eq{step1}, we write
\begin{eqnarray}
[\q,\q'] \lineup = -2\,\P(\eps,\eps')+\[\Q,-2\,\ppi(\eps,\eps')-\[\ssigma_1|_{1},\s'_1|_{-1}\] - \[\s_1|_{-1},\ssigma'_1|_{1}\]\]\nonumber\\
\lineup\ \ \ \ \ \ +\,\[\S_1|_{-1},\[\ssigma'_1|_{-1},\m_2|_2\]\]-\[\S_1'|_{-1},\[\ssigma_1|_{-1},\m_2|_2\]\]\nonumber\\
\lineup \ \ \ \ \ \ +\, \[\m_2|_2,\[\s_1|_{1},\ssigma'_1|_{-1}\]+\[\ssigma_1|_{-1},\s'_1|_{1}\]\]\nonumber\\
\lineup\ \ \ \ \ \  +\, \[\[\ssigma_1|_{-1},\m_2|_2\],\[\ssigma'_1|_{-1},\m_2|_2\]\].\label{eq:step2}
\end{eqnarray}
Here we pulled a $\Q$ out of the second and third terms in \eq{step1} and an $\m_2|_2$ out of the sixth and seventh terms in \eq{step1} using the fact that $\s_1|_1$ annihilates $\m_2|_2$. Now we can rewrite the term on the third line of \eq{step2}:
\begin{eqnarray}\[\m_2|_2,\[\s_1|_{1},\ssigma'_1|_{-1}\]+\[\ssigma_1|_{-1},\s'_1|_{1}\]\] \lineup = \[\m_2|_2,\[\s_1,\ssigma'_1\]+\[\ssigma_1,\s'_1\]-\[\s_1|_{-1},\ssigma'_1|_{1}\]-\[\ssigma_1|_{1},\s'_1|_{-1}\]\],\nonumber\\
\lineup = \[\m_2,\[\s_1,\ssigma'_1\]+\[\ssigma_1,\s'_1\]\]|_2-[\m_2|_2,\[\s_1|_{-1},\ssigma'_1|_{1}\]+\[\ssigma_1|_{1},\s'_1|_{-1}\]\],\nonumber\\
\lineup = [\m_2|_2,-2\,\ppi(\eps,\eps')-\[\s_1|_{-1},\ssigma'_1|_{1}\]-\[\ssigma_1|_{1},\s'_1|_{-1}\]\].\end{eqnarray}
Here we used the fact that $\s_1,\ssigma_1$ and $\ppi(\eps,\eps')$ are derivations of $\m_2$. Therefore \eq{step2} simplifies to
\begin{eqnarray}
[\q,\q'] \lineup = -2\,\P(\eps,\eps')+\[\Q+\m_2|_2,-2\,\ppi(\eps,\eps')-\[\s_1|_{-1},\ssigma'_1|_{1}\]-\[\ssigma_1|_{1},\s'_1|_{-1}\]\]\nonumber\\
\lineup\ \ \ \ \ \ +\,\[\S_1|_{-1},\[\ssigma'_1|_{-1},\m_2|_2\]\]-\[\S_1'|_{-1},\[\ssigma_1|_{-1},\m_2|_2\]\]\nonumber\\
\lineup\ \ \ \ \ \  +\, \[\[\ssigma_1|_{-1},\m_2|_2\],\[\ssigma'_1|_{-1},\m_2|_2\]\].\label{eq:step3}
\end{eqnarray}
Now we have to see what to do with the terms on the second two lines. For the second line note that we can rewrite
\begin{eqnarray}
\[\S_1|_{-1},\[\ssigma'_1|_{-1},\m_2|_2\]\]-\[\S_1'|_{-1},\[\ssigma_1|_{-1},\m_2|_2\]\]\lineup = \[\S_1|_{-1},\[\ssigma'_1|_{-1},\m_2|_2\]\]-\[\ssigma_1|_{-1},\[\S'_1|_{-1},\m_2|_2\]\],\nonumber\\
\lineup = [\Q,\[\ssigma_1|_{-1},\[\ssigma'_1|_{-1},\m_2|_2\]\]\].
\end{eqnarray}
and for the third line in \eq{step3}
\begin{equation}\[\[\ssigma_1|_{-1},\m_2|_2\],\[\ssigma'_1|_{-1},\m_2|_2\]\] = \[\m_2|_2,\[\ssigma_1|_{-1},\[\ssigma'_1|_{-1},\m_2|_2\]\]\],\end{equation}
using the fact that $\m_2|_2$ is nilpotent. Taken all together, this implies that the supersymmetry algebra takes the form
\begin{equation}
[\q,\q'] = -2\,\P(\eps,\eps')+\[\Q+\m_2|_2,-2\,\ppi(\eps,\eps')-\[\s_1|_{-1},\ssigma'_1|_{1}\]-\[\ssigma_1|_{1},\s'_1|_{-1}\]+\[\ssigma_1|_{-1},\[\ssigma'_1|_{-1},\m_2|_2\]\]\],
\end{equation}
and therefore the coderivation $\Ll$ in \eq{susyalg} is 
\begin{equation}\bm{\Lambda} \equiv -2\,\ppi(\eps,\eps')- \[\ssigma_1|_{1},\s'_1|_{-1}\] - \[\s_1|_{-1},\ssigma'_1|_{1}\]+\[\ssigma_1|_{-1},\[\ssigma'_1|_{-1},\m_2|_2\]\].\label{eq:Lambda}\end{equation}
Conjugating by $\G$, this implies that the supersymmetry algebra for the original field $\tilde{\Phi}$ in the small Hilbert space takes the form:
\begin{equation}\[\tilde{\S},\tilde{\S}'\] = -2\,\P(\eps,\eps') + \[\tilde{\M},\tilde{{\bm\Lambda}}\],\end{equation}
where
\begin{equation}\tilde{{\bm\Lambda}} = \G^{-1}\Ll\G\end{equation}
and we use the fact that the gauge products carry zero momentum and therefore commute with $P_\mu$. Now we have to show that $\Ll$ is consistent with the constraint on the string field. Compute:
\begin{eqnarray}
[\n,\Ll] \lineup = -2\,\p(\eps,\eps') -\[\s_1|_{1},\s'_1|_{-1}\] - \[\s_1|_{-1},\s'_1|_{1}\] +\[\s_1|_{-1},\[\ssigma'_1|_{-1},\m_2|_2\]\]-\[\ssigma_1|_{-1},\[\s'_1|_{-1},\m_2|_2\]\],\nonumber\\
\lineup = -2\,\p(\eps,\eps') -[\s_1,\s'_1]|_0 - \[\s_1|_{-1},\[\ssigma'_1|_{1},\m_2|_0\]\]-\[\ssigma_1|_{-1},\[\s'_1|_{1},\m_2|_0\]\],\nonumber\\
\lineup = -\[\[\s_1|_{-1},\ssigma'_1|_{1}\],\m_2|_0\]-\[\ssigma'_1|_{1},\[\s_1|_{-1},\m_2|_0\]\]-\[\[\ssigma_1|_{-1},\s'_1|_{1}\],\m_2|_0\]-\[\s'_1|_{1},\[\ssigma_1|_{-1},\m_2|_0\]\],\nonumber\\
\lineup = \[\m_2|_0,\[\s'_1|_{1},\ssigma_1|_{-1}\]+\[\ssigma'_1|_{1},\s_1|_{-1}\]\]-\[\ssigma'_1|_{1},\[\s_1,\m_2\]|_{-1}\]-\[\s'_1|_{1},\[\ssigma_1,\m_2\]|_{-1}\],\nonumber\\
\lineup = \[\m_2|_0,-2\,\ppi(\eps,\eps')+ \[\s'_1|_{-1},\ssigma_1|_{1}\] + \[\ssigma'_1|_{1}\s_1|_{-1}\]+\[\ssigma_1|_{-1},\[\ssigma'_1|_{-1},\m_2|_2\]\]\],\nonumber\\
\lineup = \[\m_2|_0,\Ll\].
\end{eqnarray}
In the first step we computed the action of $\n$ on $\Ll$; in the second step we used conservation of Ramond number in addition to the derivation property of $\sigma_1$ and $s_1$; in the third step we canceled the first two terms using \eq{ssp} and used the Jacobi identity; in the fourth step we collected terms and used Ramond number conservation; in the fifth step we used the derivation property of $s_1$ and $\sigma_1$; in the sixth step we added terms which vanish because of associativity of $m_2$ and the derivation property of $s_1$ and $\sigma_1$; in the sixth step we substituted the definition of $\Ll$. Note that we needed to express part of the translation operator in the form $[Q,\pi(\eps,\eps')]$ so that $\Ll$ would be consistent with this constraint.

So far we have been working at the level of coderivations on the tensor algebra, but it is helpful to bring things down to earth and express the supersymmetry algebra in terms of the NS and R string fields $\varphi_\mathrm{N}$ and $\psi_\mathrm{R}$. Acting equation \eq{susyalg} on the group-like element of $\tilde{\varphi}$, projecting onto the 1-string component of the tensor algebra, and separating NS and R components produces an expression of the form
\begin{eqnarray}
[\delta,\delta']\varphi_\mathrm{N} \lineup = -2\,P(\eps,\eps') \varphi_\mathrm{N}+\left({ \mathrm{gauge} \atop \mathrm{transformation}}\right)_\mathrm{N} + \left({ \mathrm{on\ shell} \atop \mathrm{trivial}}\right)_\mathrm{N},\label{eq:nontsusy1}\\
\ [\delta,\delta']\psi_\mathrm{R} \lineup = -2\,P(\eps,\eps') \psi_\mathrm{R}+\left({ \mathrm{gauge} \atop \mathrm{transformation}}\right)_\mathrm{R} + \left({ \mathrm{on\ shell} \atop \mathrm{trivial}}\right)_\mathrm{R}.\label{eq:nontsusy2}
\end{eqnarray}
The extra terms represent an infinitesimal gauge transformation and a piece which vanishes assuming the equations of motion. An infinitesimal gauge transformation of the polynomial equations of motion \eq{poly} takes the general form
\begin{eqnarray}
\left({ \mathrm{gauge} \atop \mathrm{transformation}}\right)_\mathrm{N}\lineup = Q\lambda_\mathrm{N} + [\psi_\mathrm{R},\lambda_\mathrm{R}],\\
\left({ \mathrm{gauge} \atop \mathrm{transformation}}\right)_\mathrm{R}\lineup =  Q\lambda_\mathrm{R},
\end{eqnarray}
where the NS and R gauge parameters $\lambda_\mathrm{N}$ and $\lambda_\mathrm{R}$ are subject to the constraints
\begin{eqnarray}
\eta\lambda_\mathrm{N}\lineup = [\varphi_\mathrm{N},\lambda_\mathrm{N}],\\
\eta\lambda_\mathrm{R}\lineup = [\varphi_\mathrm{N},\lambda_\mathrm{R}]+[\psi_\mathrm{R},\lambda_\mathrm{N}].
\end{eqnarray}
The particular gauge parameters which appear the supersymmetry algebra are
\begin{eqnarray}
\lambda_\mathrm{N} \lineup \equiv -2\,\pi(\eps,\eps') \varphi_\mathrm{N} -(\sigma_1s_1'-\sigma_1's_1)\varphi_\mathrm{N}-[\sigma_1\varphi_\mathrm{N},\sigma_1'\varphi_\mathrm{N}],\\
\lambda_\mathrm{R} \lineup \equiv  -2\,\pi(\eps,\eps') \psi_\mathrm{R} -(s_1\sigma_1'-s_1'\sigma_1)\psi_\mathrm{R}+\sigma_1[\psi_\mathrm{R},\sigma_1'\varphi_\mathrm{N}]-\sigma_1'[\psi_\mathrm{R},\sigma_1\varphi_\mathrm{N}].
\end{eqnarray}
One can check that the gauge parameters satisfy the constraints, which is basically a consequence of the fact that $\Ll$ satisfies \eq{Lconst}. Now let's write down the on-shell trivial terms. For short, let us write the NS and Ramond Euler-Lagrange functions
\begin{eqnarray}
E_\mathrm{N}\lineup \equiv Q\varphi_\mathrm{N} +\psi_\mathrm{R}*\psi_\mathrm{R},\\
E_\mathrm{R}\lineup \equiv Q\psi_\mathrm{R}.
\end{eqnarray}
Then
\begin{eqnarray}
\left({ \mathrm{on\ shell} \atop \mathrm{trivial}}\right)_\mathrm{N}\lineup = -2\,\pi(\eps,\eps') E_\mathrm{N} -(\sigma_1s_1'-\sigma_1's_1)E_\mathrm{N}+[\sigma_1'\varphi_\mathrm{N},\sigma_1E_\mathrm{N}]-[\sigma_1\varphi_\mathrm{N},\sigma_1'E_\mathrm{N}],\\
\left({ \mathrm{on\ shell} \atop \mathrm{trivial}}\right)_\mathrm{R}\lineup = -2\,\pi(\eps,\eps') E_\mathrm{R} -(s_1\sigma_1'-s_1'\sigma_1)E_\mathrm{R}+\sigma_1[\sigma_1'E_\mathrm{N},\psi_\mathrm{R}]+\sigma_1[E_\mathrm{R},\sigma_1'\varphi_\mathrm{N}]
\nonumber\\
\lineup\ \ \ \ \ \ \ \ \ \ \ \ \ \ \ \ \ \ \ \ \ \ \ \ \ \ \ \ \ \ \ \ \ \ \ \ \ \ \ \ \ \ \ \ \ \ \, 
-\,\sigma'_1[\sigma_1E_\mathrm{N},\psi_\mathrm{R}]-\sigma_1'[E_\mathrm{R},\sigma_1\varphi_\mathrm{N}].
\end{eqnarray}
Once we impose the equations of motion, we obtain a supersymmetry algebra of the expected form modulo gauge transformations.

\section{Conclusions}

In this paper we have constructed consistent classical field equations for all superstring theories, and for the open superstring given an explicit analysis of supersymmetry. A proof that our field equations imply the correct tree-level amplitudes will be provided in upcoming work \cite{Konopka}. Let us conclude by discussing future directions. 

Though we don't know how to write a fully satisfactory action for the Ramond sector, it should be possible to formulate a tree-level action which includes two Ramond string fields (typically, at picture $-1/2$ and $-3/2$), which are afterwards related by imposing a ``self-dual" constraint on classical solution space \cite{Michishita}. See \cite{Kunitomodual} for recent discussion. One version of this idea was recently suggested in \cite{1PIR}, and would be particularly natural to implement using the methods of this paper. However, the required products in the equations of motion will be different from those introduced here; in a sense they will be more complicated, since even at a given Ramond number the products will differ depending on the number of Ramond states being multiplied.  However, this is probably not an insurmountable complication. It would be particularly nice if an action with constraint could be realized for type II closed superstring field theory, as it would give a potentially interesting gauge invariant observable for Ramond-Ramond backgrounds. However, it remains to be seen whether an action with constraint helps in defining the quantum theory.

One important question is whether recent developments in superstring field theory can help in understanding higher genus amplitudes in superstring perturbation theory. The conservative approach to this problem requires first constructing a satisfactory classical action, and then quantizing following the methodology of the Batalin-Vilkovisky formalism \cite{Zwiebach}. However, given present limitations in the Ramond sector, a more pragmatic approach may be to construct a 1PI effective superstring field theory, as suggested by Sen \cite{1PI,1PIR}. The main question in this respect is whether the methods developed here and in previous works can be adapted to handle spurious singularities which appear in superconformal ghost correlators at higher genus \cite{Verlinde}. It may also be helpful to clarify the relation between our construction of vertices and the method of ``vertical integration" introduced in \cite{SenOS} and further developed in \cite{SenWitten}. We hope to return to these questions soon.

\vspace{.5cm}

\noindent{\bf Acknowledgments}

\vspace{.25cm}

\noindent T.E. would like to thank S. Hellerman and A. Sen for conversations. This work was supported in parts by the DFG Transregional Collaborative Research Centre TRR 33 and the DFG cluster of excellence Origin and Structure of the Universe.

\end{document}